\documentclass{jfm}

\usepackage{graphicx}
\usepackage{natbib}

\ifCUPmtlplainloaded \else
  \checkfont{eurm10}
  \iffontfound
    \IfFileExists{upmath.sty}
      {\typeout{^^JFound AMS Euler Roman fonts on the system,
                   using the 'upmath' package.^^J}%
       \usepackage{upmath}}
      {\typeout{^^JFound AMS Euler Roman fonts on the system, but you
                   dont seem to have the}%
       \typeout{'upmath' package installed. JFM.cls can take advantage
                 of these fonts,^^Jif you use 'upmath' package.^^J}%
      }
  \else
  \fi
\fi

\ifCUPmtlplainloaded \else
  \checkfont{msam10}
  \iffontfound
    \IfFileExists{amssymb.sty}
      {\typeout{^^JFound AMS Symbol fonts on the system, using the
                'amssymb' package.^^J}%
       \usepackage{amssymb}%
         \let\leq=\leqslant
         \let\geq=\geqslant
      }{}
  \fi
\fi

\ifCUPmtlplainloaded \else
  \IfFileExists{amsbsy.sty}
    {\typeout{^^JFound the 'amsbsy' package on the system, using it.^^J}%
     \usepackage{amsbsy}}
    {}
\fi

\usepackage{tikz}
\usetikzlibrary{decorations.pathmorphing}

\usepackage{time}

\usepackage{graphicx}
\usepackage{dcolumn}

\def\ie{{i.e.}~}
\def\cf{{cf.}~}

\newcommand\eg{e.g.~}

\def\fig#1{\ref{fig:#1}}
\def\Fig#1{figure~\fig{#1}}
\def\Figs#1{figures~\fig{#1}}
\def\FIG#1{Figure~\fig{#1}}

\newcommand\tableRef[1]{\ref{tab:#1}}
\newcommand\Tab[1]{Table~\tableRef{#1}}

\def\eq#1{(\ref{eq:#1})}
\def\Eq#1{\eq{#1}}
\def\EQ#1{Equation~\eq{#1}}

\def\Sect#1{section~\ref{sec:#1}}
\def\App#1{appendix~\ref{sec:#1}}

\def\rmd{\ensuremath{\mathrm{d}}}
\def\dichte{\ensuremath{n}}
\def\Rav{\ensuremath{\langle a \rangle}}
\def\collEfficiency{\ensuremath{\varepsilon}}

\def\visc{\ensuremath{\mu}}

\newcommand\Pen{{\textit{Pe}}}  

\title{Crossing the bottleneck of rain formation}

\author[M. Rohloff, T. Lapp and J. Vollmer]%
{M. Rohloff, T. Lapp and J. Vollmer\thanks{Email address for correspondence: \texttt{juergen.vollmer@ds.mpg.de}}}
\affiliation{{Max Planck Institute for Dynamics and Self-Organization (MPI DS), 37077 G\"ottingen, Germany}\\[\affilskip]
{Faculty of Physics, Georg-August Universit\"at~G\"ottingen, 37077 G\"ottingen, Germany}} 

\pubyear{}
\volume{}
\pagerange{}

\date{\today -- \now}

\begin{document}

\maketitle

\begin{abstract}
  The demixing of a binary fluid mixture, under gravity, is a two
  stage process.  Initially droplets, or in general aggregates, grow
  diffusively by collecting supersaturation from the bulk phase.
  Subsequently, when the droplets have grown to a size, where their
  P\`eclet number is of order unity, buoyancy substantially enhances
  droplet growth. The dynamics approaches a finite-time singularity
  where the droplets are removed from the system by precipitation.
  The two growth regimes are separated by a bottleneck of minimal
  droplet growth.
  Here, we present a low-dimensional model addressing the time span
  required to cross the bottleneck, and we hence determine the time,
  $\Delta t$, from initial droplet growth to rainfall.
  Our prediction faithfully captures the dependence of $\Delta t$ on
  the ramp rate of the droplet volume fraction, $\xi$, the droplet
  number density, the interfacial tension, the mass diffusion
  coefficient, the mass density contrast of the coexisting phases, and
  the viscosity of the bulk phase.  The agreement of observations and
  the prediction is demonstrated for methanol/hexane and
  isobutoxyethanol/water mixtures where we determined $\Delta t$ for a
  vast range of ramp rates, $\xi$, and temperatures.
  The very good quantitative agreement demonstrates that it is sufficient
  for binary mixtures to consider (i) droplet growth by
  diffusive accretion that relaxes supersaturation, and (ii) growth by
  collisions of sedimenting droplets. 
  An analytical solution of the resulting model provides a
  quantitative description of the dependence of $\Delta t$ on the ramp
  rate and the material constants.
  Extensions of the model that will admit a quantitative prediction of
  $\Delta t$ in other settings are addressed.
\end{abstract}


\section{Introduction}

Precipitation emerges when aggregates, \ie droplets, bubbles or solid
particles that are immersed in a fluid, grow to a size where their motion is affected by buoyancy.
At this point their motion changes from
Brownian diffusion to Stokes sett\-ling, and the collision cross section
increases dramatically. As a consequence 
aggregate growth is boosted \citep{Houghton1959,GrawLiu2003,GrabowskiWang2013}, 
collective effects emerge in their motion \citep{CauLacelle1993,KalwarczykZiebaczFialkowskiHolyst2008,StevensFeingold2009,Woods2010}, and
virtually all volume condensed on the aggregates is precipitating out
of the fluid in a finite time \citep{CauLacelle1993,AartsDullensLekkerkerker2005,KostinskiShaw2005}. 
Precipitation is prevalent in natural processes, such as 
clouds
\citep{Houghton1959,GrawLiu2003,StevensFeingold2009,Tokano2011}, hot-
\citep{IngebritsenRojstaczer1993,ToramaruMaeda2013} and cold-water
\citep{HanLuMcPhersonKeatingMooreEtAl2013} geysers, as well as lake
\citep{Zhang1996,ZhangKling2006} and volcano
\citep{WylieVoightWhitehead1999,CashmanSparks2013} eruptions and the
subsequent cooling of magma domes
\citep{MartinNokes1988,KoyaguchiHallworthHuppertSparks1990,SparksHuppertKozaguchiHallwood1993}.
Moreover, it is also essential to many technical processes, like
synthesis of large colloidal particles
\citep{NozawaDelvilleUshikiPanizzaDelville2005}, steel processing
\citep{YuanThomasVanka2004,RimbertClaudotteGardinLehmann2014}, and
food science \citep{ScholtenLindenThis2008,ZhangXu2008}.

\begin{figure}
\[ \includegraphics[width=0.7\textwidth]{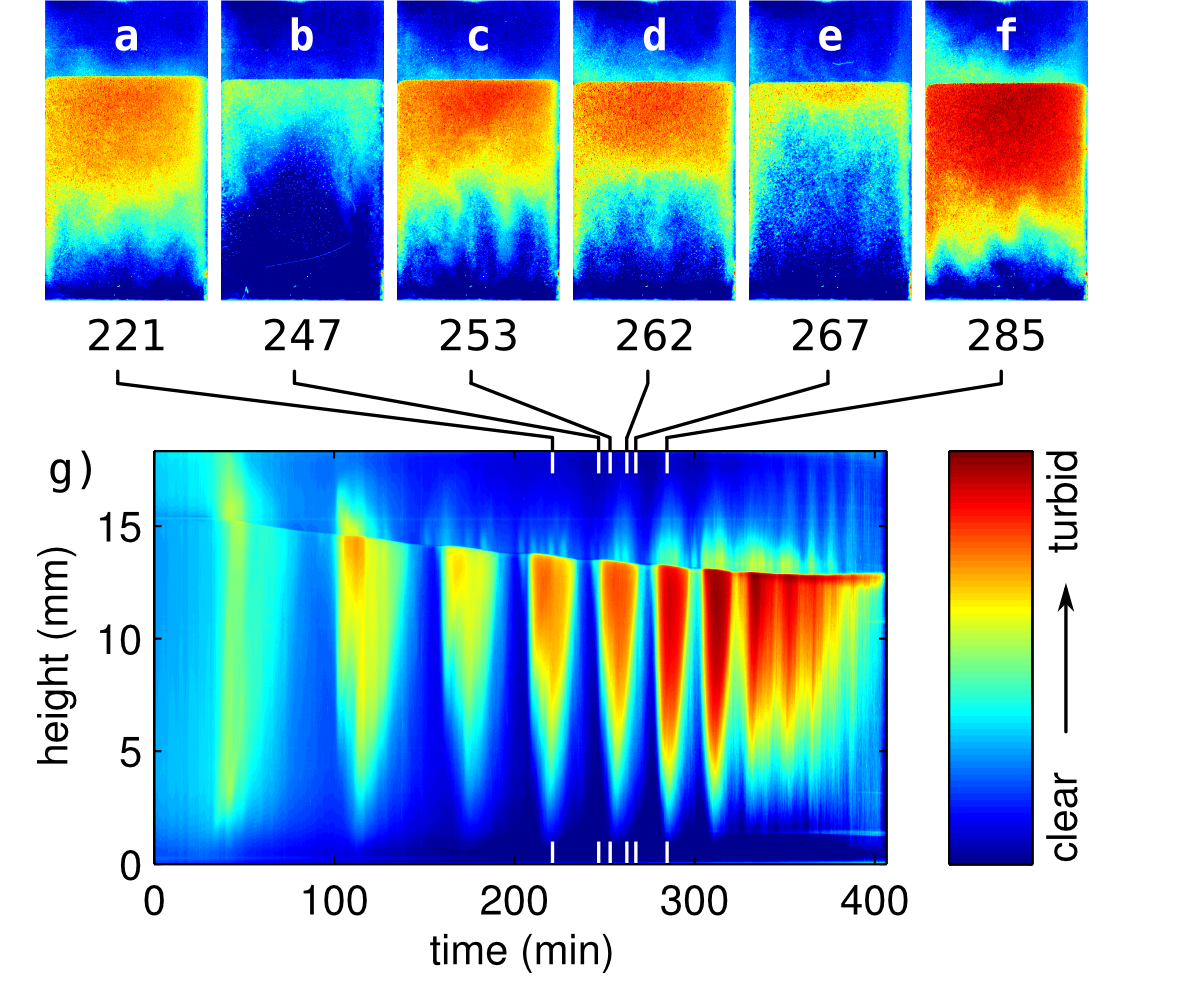} \]
\caption[]{\textbf{Episodic precipitation in binary mixtures.} 
  Panels a--f show false-colour plots of the turbidity distribution in 
  snap shots of the phase separation of an
  isobutoxyethanol/water mixture subjected to a ramp rate, 
  $\xi=2.5\times 10^{-5}\,$s$^{-1}$.  Averaging in horizontal direction and
  arranging the resulting vertical turbidity profiles next to each
  other produces a space-time plot of the time evolution of the
  turbidity, panel g. The length scale is provided on its ordinate axis, and 
  the scales in the pictures a)--f) can be inferred by noticing that panel g shows 
  the full height of the samples. 
  In \Sect{experimental} we provide full details on the experimental 
  setup and method, and in Movie~1 we supply the full evolution in the false colours 
presented here, together with an animation illustrating the construction of the space-time plot. 
\label{fig:spacetime}}
\end{figure}

In spite of the abundance of applications, there are many unresolved
issues in the quantitative description of precipitation.  
For instance, a better understanding of rain formation has been
identified as one of the key ingredients of improved models for
climate modeling \citep{GrabowskiWang2013,BlythLowensteinHuangCuiDaviesCarslaw2013} and 
small-scale weather prediction \citep{StevensSeifert2008}.
Here, we present a comprehensive set of experimental data that allows
us to critically survey the parameter dependence of the time scale
$\Delta t$ for rain formation.
Data are provided for two binary fluid mixtures where demixing is driven by a continuous 
temperature ramp. 
When the ramp induces a constant
generation of material, characterised by a constant value of the ramp rate, $\xi$,
we observe
repeated waves of aggregate nucleation, growth and
precipitation (\Fig{spacetime}). We denote these waves of
precipitation as \emph{episodic precipitation}.
The defining feature of episodic precipitation is an oscillatory
evolution of the aggregate size distribution and of the precipitation
rate in response to a slow continuous mass or heat flux into a fluid
mixture. The flux leads to aggregate nucleation and growth, and
episodic release of the accumulated material by precipitation
events. The modulations of the precipitation rates has been observed
in laboratory experiments where phase separation in a binary fluid was
monitored during pressure release
\citep{SoltzbergBowersHofstetter1997} or a temperature ramp
\citep{MirzaevHeimburgKaatze2010,auernhammer05JCP,LappRohloffVollmerHof2012}.

Constant driving, $\xi$, induces periodic waves of precipitation in
both coexisting phases (\Fig{spacetime}).  We identify the time scale
$\Delta t$ for rain formation as the period of the episodic response
in the observed demixing.  Hence, we obtain comprehensive data sets
for the dependence of the time scale $\Delta t$ on the viscosity, the
diffusion coefficient, the mass density contrast, the number density
of aggregates and the driving.  The latter all vary over several
orders of magnitude in our experiments (\cf \App{MatConst}).

The data on $\Delta t$ is compared to a low-dimensional model that accounts for diffusive growth of
small aggregates, and a crossover to collection-dominated growth for
large aggregates.  
The model differs from classical models of rain formation by
modeling the diffusive growth according to state-of-the-art models for
nanoparticle synthesis
\citep{Sugimoto1992,TokuyamaEnomoto1993,Leubner2000,ClarkKumarOwenChan2011},
rather than adapting classical Ostwald ripening
\citep{Houghton1959,Wilkinson2014}.  We will show that these
assumptions are sufficient to quantitatively predict the
values of $\Delta t$ for the demixing of binary fluid mixtures, and to
faithfully capture the dependence of the period on their material
constants, the number density of droplets and the ramp rate.

For the demixing of binary fluid mixtures the time scale, $\Delta t$,
is selected by a bottleneck arising at the crossover from the
diffusive growth of small aggregates to growth dominated by collection
of other aggregates.  The crossover emerges once the motion of the
largest aggregates is affected by buoyancy.
All applications mentioned above share conditions where the overall
droplet volume is growing in time. Under these conditions the
diffusive growth is typically dramatically faster than for classical
Ostwald ripening, \ie in circumstances where the overall droplet
volume is preserved and the droplet number decays like one over time.
Indeed, for all experimentally accessible ramp rates, $\xi > 0$,
droplet growth progresses at a constant aggregate number density
\citep{Sugimoto1992,Leubner2000,TokuyamaEnomoto1993,ClarkKumarOwenChan2011,VollmerPapkeRohloff2014}.
The focus of the present paper will therefore be the characterisation
and modeling of aggregate growth and precipitation in settings with a
sustained constant growth speed $\xi$ of the overall aggregate volume
fraction, and the analysis of the dependence of $\Delta t$ on $\xi$,
the aggregate concentration $\dichte$, and appropriate material
constants.

The paper is organised as follows: 
In \Sect{experimental} we provide details on the considered
mixtures, and the experimental procedure to determine $\Delta t$. 
It culminates in the presentation of a large data set that clearly 
establishes a strong dependence of $\Delta t$ on the ramp rate~$\xi$. 
The robust features of episodic precipitation call for a universal
description of the oscillation period.  Such a theory is established
in \Sect{theory}.  The resulting prediction is in very good
quantitative agreement with the experimental data.  (All material
constants needed for the quantitative comparison are provided in
\App{MatConst}.)
The model allows us to revisit problems encountered in quantitative
descriptions of warm terrestrial rain (\Sect{discussion}): diffusive
droplet growth in a classical Ostwald-like scenario is too slow to
account for the observed time scale, $\Delta t$.  In contrast, our new
model provides estimates for clouds that are too fast.  We attribute
this to simplifications of the droplet collision kernel that are
well-justified for binary mixtures with relatively small settling
rates, but that substantially overestimate the growth rate in systems,
like terrestrial rain, with large density contrast of the coexisting
phases.
We conclude in \Sect{conclusion} with a summary of our main results,
and a discussion of extensions of the model that will allow us to address 
precipitation arising in other settings.

\section{Experiment}
\label{sec:experimental}

\begin{figure}
\[ \includegraphics[width=0.7\textwidth]{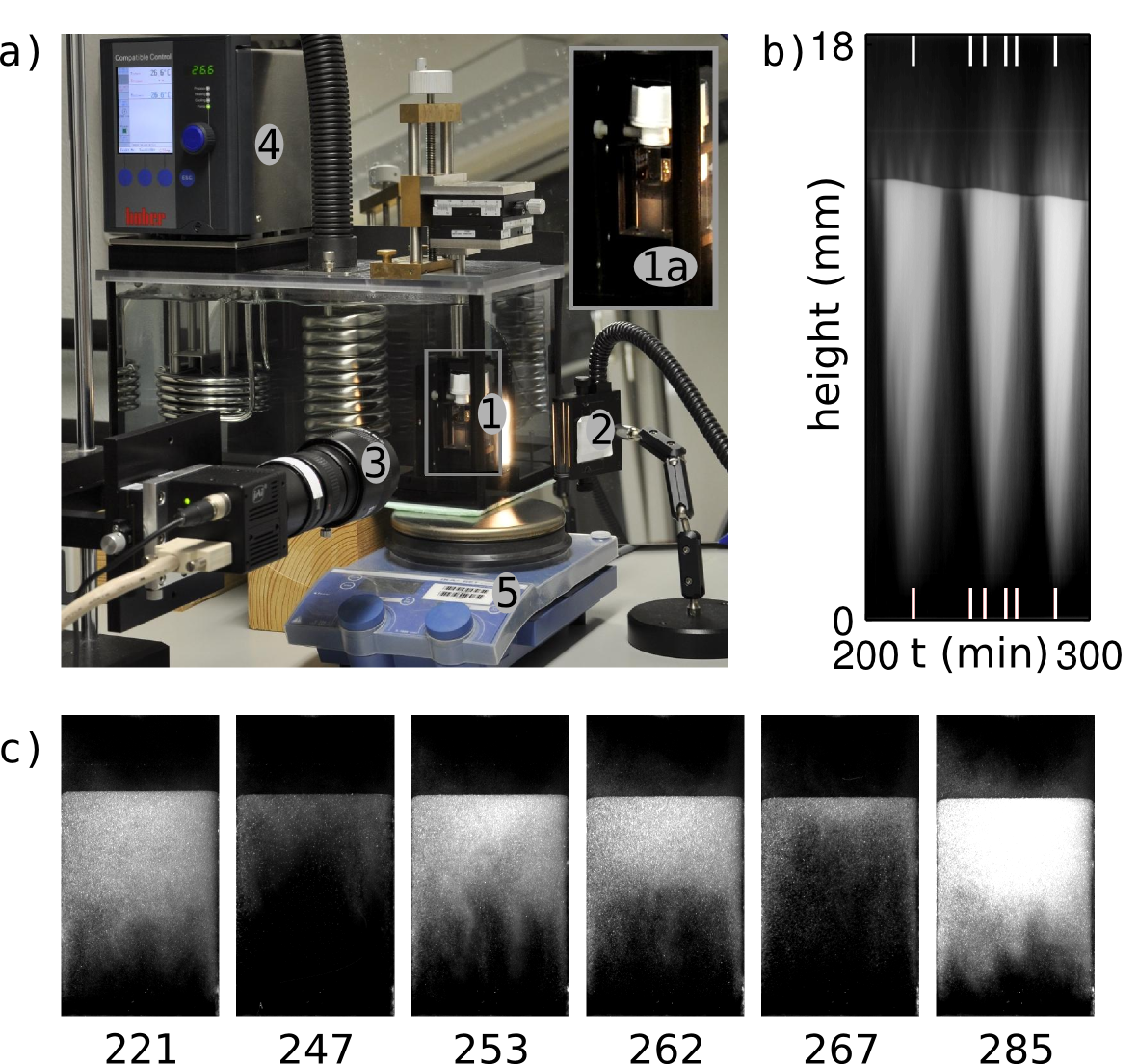} \]
\caption[]{\textbf{Setup and raw data obtained from the CCD camera.}
  a)~Photo of the experimental setup. Its components, (1)--(5) are described in the main text.
  b)~Space-time plot based on the turbidity data provided by the camera. 
  The sample contains a mass fraction of $0.2751$ isobutoxyethanol in water, and was subjected to a ramp rate~$\xi=2.5\times 10^{-5}\,$s$^{-1}$. 
  c)~Representative snap shots of the times where the snap shots were taken are indicated below the respective photos, and by
  white bars in panel b).
  In Movie~2 we provide the full time evolution as captured by our camera. 
  \label{fig:setup}}
\end{figure}

We will discuss the parameter dependence of $\Delta t$ for repeated
waves of precipitation in two well-controlled laboratory experiments:
mixtures of isobutoxyethanol/water and of methanol/hexane that are
subjected to a range of different temperature ramps. The system is
contained in a light scattering cuvette and its temperature is
controlled by immersion in a water bath so that we have full control
over external perturbations.
In our experiment, \Fig{spacetime}, two partially miscible liquids
form two layers with a phase which is richer in the less dense fluid
floating over a layer of the high-density phase.  
The temperature of the mixture is varied smoothly away from
the phase coalescence point,~$T_{\rm c}$, and the time-dependence of the
temperature is engineered so that the ramp rate, $\xi$, of the droplet 
volume fraction remains constant in each run of the experiment. 
A movie illustrating the corresponding temperature evolution together
with a video of the sample is provided in movie~2.
In response to the ramp both layers show an alternating variation in
turbidity, \Fig{spacetime}.a)--f) and \Fig{setup}.c).
Representing this evolution in a space-time plot, \Fig{spacetime}.g),
illustrates a variation of turbidity with a period $\Delta t_i$
between the $i^{\textrm{th}}$ and $(i+1)^{\textrm{st}}$ precipitation
event.
The accompanying periodic alternation in the turbidity and the
particle size distribution are characteristics of episodic
precipitation.  The effect is robust. Episodic response has been
observed in the particle size distribution
\citep{LappRohloffVollmerHof2012} and in calorimetric data
\citep{vollmer97JCP1, vollmer99, auernhammer05JCP,
  MirzaevHeimburgKaatze2010} in a vast range of binary mixtures
\citep{vollmer97JCP1,auernhammer05JCP,MirzaevHeimburgKaatze2010,LappRohloffVollmerHof2012},
including olive oil and methylated spirit \citep{vollmer07PRL}.  It
arises in the upper as well as in the lower layer of the mixtures.

\subsection{Experimental Setup}
\label{sec:setup}

\FIG{setup}.a) shows the experimental setup.  The sample cell~(1), a
$3\;$mL fluorescence cell 117.100F-QS made by Hellma GmbH, is
illuminated by a KL 2500 LCD Schott cold light source~(2) such that
dark-field images can be taken with a BM-500CL monochrome progressive
scan CCD camera~(3).  The camera takes $772 \times 1420$ pixel images
of the sample cell with a frame rate between $0.1$ and $3\;$Hz
depending on the ramp rate~$\xi$.

The sample temperature is controlled by immersion into a water bath
that follows a temperature protocol imposed by a computer-controlled
thermostat~(4): an immersion cooler Haake EK20 is cooling with
constant power, and a Huber \hbox{CC-E} immersion thermostat is heating the
water bath to the preset temperature. Additionally, the temperature of
the water near the sample is measured with a PT100 temperature
sensor. The temperature is controlled with an accuracy of $15\,$mK.
Homogenisation for repeated runs is provided by a
magnetic stirring unit~(5).

The inset~(1a) in \Fig{setup}.a) shows a magnification of the sample cell.
The camera captures the turbidity of the full cell, providing 8 bit
turbidity data as shown in \Fig{setup}.c). Averaging this data in
horizontal direction and plotting the resulting scans of the turbidity
height profiles, provides the space-time plot \Fig{setup}.b).
For visual inspection the contrast in these pictures is conveniently
enhanced by a representation in false colours, \Fig{spacetime}.
As supplementary online material we provide movies showing the
black-and-white turbidity data taken by the camera together with a
plot of the time evolution of the temperature, Movie~2, and an animation, Movie~1,
illustrating the construction of the space-time plot of the turbidity
shown in \Fig{spacetime}.g).

The space-time plots, \Figs{spacetime}.g) and \fig{setup}.c), clearly visualise the period
$\Delta t$ between subsequent waves of precipitation.
Episodic precipitation goes along with marked oscillatory changes in the
droplet size distribution \citep{LappRohloffVollmerHof2012}, and in the
turbidity of the samples \citep{auernhammer05JCP}. The main panels of
\Fig{phasdias} show representative traces of the turbidity of the
samples when heated with different, constant~$\xi$.
The data for $\Delta t$ are extracted from these traces as the distance between subsequent maxima of the turbidity.

In addition to capturing the turbidity we succeeded to follow the time
evolution of the droplet size distribution of IBE droplets in water
via an appropriately enhanced illumination and imaging
\citep{LappRohloffVollmerHof2012}.  Wherever available an analysis of
the temporal evolution of the droplet size distributions along the
same line as the one for the space-time plots of the turbidity,
provides identical values for $\Delta t_i$ with a higher experimental
accuracy.  Further details on the experimental setup are provided in
\citet{LappRohloffVollmerHof2012}, and the data analysis used to
extract the oscillation period from the space-time plots has been
described in \citet{auernhammer05JCP}.

\subsection{Investigated Mixtures}
\label{sec:phasdia}

\begin{figure}
\includegraphics[width=\textwidth]{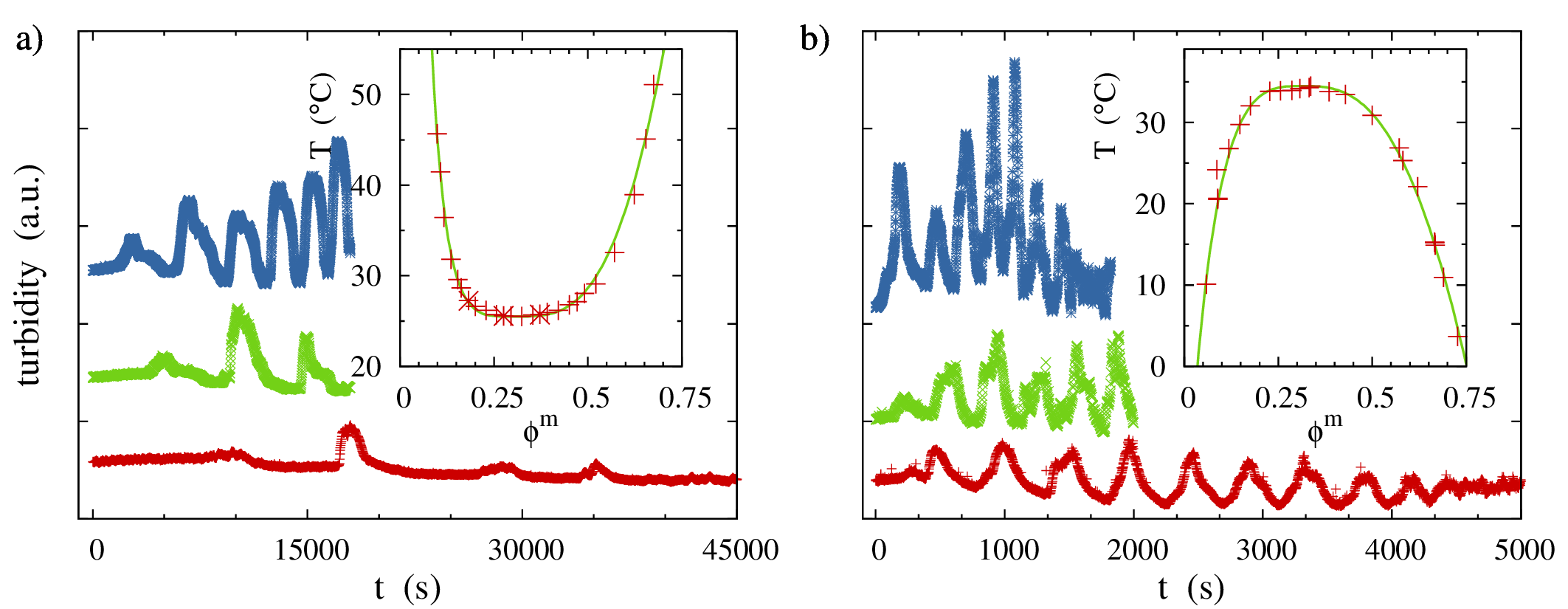} 
\caption[]{\textbf{Phase diagrams and time traces of the turbidity.}
  The insets show the phase diagrams of a) IBE+W, and b) M+H, and the
  respective main panels show oscillations of the turbidity in the
  lower layer for different ramp rates~$\xi$: from top to bottom a)
  $2.5 \times 10^{-5}$s$^{-1}$, $1.25 \times 10^{-5}$s$^{-1}$, $5
  \times 10^{-6}$s$^{-1}$ for IBE+W, and b) $4 \times
  10^{-4}$s$^{-1}$, $2 \times 10^{-4}$s$^{-1}$, $1 \times
  10^{-4}$s$^{-1}$ for M+H.  The data points of the turbidity
  correspond to the average turbidity of a region of $140\,\mu$m
  height and the full sample width, that is located $650\,\mu$m below
  the meniscus.  The different signals are shifted vertically for
  better visibility.
  \label{fig:phasdias}
}
\end{figure}

Two types of mixtures are considered:
\begin{description}
\item[Methanol/hexane (M+H) \ ] These mixtures are one of the classical
  model systems of binary phase
  separation \citep{huang74,BeysensGuenounPerrot1988,AbbasSatherleyPenfold1997,IwanowskiSattarowBehrendsMirzaevKaatze2006,SamAuernhammerVollmer2011}.
  The two liquids are fully miscible above the critical temperature
  $T_{\rm c}=34.45^\circ {\rm C}$.  The concentrations of the coexisting phases
  that are formed for lower temperatures are shown in the phase
  diagram in \Fig{phasdias}.b).  See \citet{AbbasSatherleyPenfold1997} for a detailed
  description.

\item[Isobutoxyethanol/water (IBE+W) \ ] Mixtures of water and
  butoxyethanol have become popular as an experimentally-friendly
  system that phase separates upon heating \citep[see \eg][]{EmmanuelBerkowitz2006}. 
  For our present purposes IBE and water, \Fig{phasdias}.a), is even preferential
  since the critical point of the mixtures, $T_{\rm c}=25.61^\circ
  {\rm C}$, lies more than $10^\circ {\rm C}$ below the one of the
  butoxyethanol mixture.  This further enhances the range of
  experimentally accessible temperatures (that must always lie well
  below the boiling point of water).  See
  \citet{NakataDobashiKuwaharaKaneko1982} and \citet{LappRohloffVollmerHof2012} for more detailed descriptions.
\end{description}

For the fit of the coexistence curve we follow the procedure of 
\citet{AizpiriCorreaRubioPena1990}.  To first order they approximate the
left and right branch of the coexistence curve by 
\begin{equation}
  \Phi_{r/l} = \Phi_c \pm B \: \theta^{\beta} + D \: \theta^{2\beta}
\end{equation}
with the reduced temperature $\theta = |1-T/T_c|$, the critical point
being at temperature $T_c$ with composition $\Phi_c$, and the universal
scaling exponent $\beta = 0.325$. 
For the M+H mixture
this provides a good fit, the solid
green line, shown in the inset of \Fig{phasdias}.b), with fit parameters listed in \Tab{phasediagram}. On
the other hand for IBE+W the exponent $\beta = 0.325$ only applies for
$\theta < 10^{-3}$ \citep{NakataDobashiKuwaharaKaneko1982}, which is
too small for our purposes. Even correction terms based on the Wegner
expansion do not help \citep{NakataDobashiKuwaharaKaneko1982}. To have
a simple set of parameters we therefore choose $\beta = 0.25$, which
admits a faithful description based on three free non trivial
parameters (see \Fig{phasdias}.a) and \Tab{phasediagram}).

\begin{table}
\centering
 \begin{tabular}{l@{$\qquad$}rcl@{$\qquad$}rcl}
			& IBE&+&W 		& 	M&+&H	\\[2mm]
    $\beta$		  & $0.25$ &&  		            & $0.325$		\\
    $T_c$ [K]   & $298.76$&$\pm$&$0.12$  	  & $307.88$&$\pm$&$0.15$ 	\\
    $\Phi_c$ 		& $0.3093$&$\pm$&$0.0032$ 	& $0.3143$&$\pm$&$0.0008$	\\
    $B$      		& $0.547$&$\pm$&$0.002$  	  & $0.726$ &$\pm$&$0.002$ 	\\
    $D$      		& $0.26$&$\pm$&$0.015$	    & $0.323$ &$\pm$&$0.005$	
  \end{tabular} 
\caption{Fit parameters of the coexistence curve for IBE+W and M+H.}
\label{tab:phasediagram}
\end{table}

The temperature ramps in our experiments amount to increasing
temperature for IBE+W, and decreasing
temperature for M+H. For simplicity we denote this as
heating, and understand that the temperature ramp rate is negative for
the latter mixture.  
On the other hand, the ramp rate of the droplet volume fraction, $\xi$, is positive in either case, as elaborated in \Sect{ramp}.

The evolution can most conveniently be described by focusing on a
region in one of the macroscopic phases. Its average concentration
changes due to sedimentation of large droplets.  However, immediately
after a precipitation event, the bulk and the remaining small droplets
are very close to an equilibrium composition at points on the coexistence curve with composition
$\Phi_b$ for the bulk phase, and $\Phi_d$ for the
remaining droplets in the fluid. The phases occupy the volumes
$V_b$ and $V_d$, respectively.

As the mixture is further heated, the equilibrium concentrations of
the coexisting phases change in response to the broadening of the
miscibility gap, \ie the region bounded by the coexistence curve.
A temperature difference $\delta T$ causes a change in the equilibrium
composition by $\delta \Phi_b<0$ and $\delta \Phi_d>0$. It
gives rise to a concentration current across the interface of the
droplets, which in turn leads to a growth of the droplets.  In the
following subsection we review how the temperature protocol of the
experiments was chosen in order to fix the ramp rate, $\xi$, of the droplet
volume fraction.

\subsection{Calculating the ramp rate $\xi$}
\label{sec:ramp}

The derivation of the ramp rate, $\xi$, starts from the the average composition 
\begin{equation}
   \phi  = v_d \Phi_d + (1-v_d) \: \Phi_b
\end{equation}
of a small volume of a mixture, where droplets of composition $\Phi_d$
occupy a volume fraction $v_d$ in a background phase of composition
$\Phi_b$. By definition, the average composition, $\phi$, is preserved when
the droplets start growing in response to a change of temperature. On
the other hand droplet growth is accompanied by a change of the
composition of the phases,
\begin{equation}
  0 = \dot \phi 
    =    v_d  \, \dot\Phi_d + \Phi_d \, \dot v_d  
    + (1-v_d) \, \dot\Phi_b - \Phi_b \, \dot v_d  \, .
\label{eq:mass-conservation}
\end{equation}
We introduce the notations
\begin{subequations}
\begin{eqnarray}
  \zeta &=& \Phi_0^{-1} \; \frac{\rmd \bar\Phi}{\rmd t}
  \\[2mm]
  \xi &=& \Phi_0^{-1} \; \frac{\rmd \Phi_0}{\rmd t}
  \\[2mm]
  \varphi &=& \frac{\phi - \bar\Phi}{\Phi_0}
  \label{eq:varphi}
  \\[2mm]
  \textrm{where} \qquad
  \bar\Phi &=& \frac{1}{2} \; \left( \Phi_b + \Phi_d \right) \, ,
  \\[2mm]
  \Phi_0 &=& \frac{1}{2} \; \left( \Phi_b - \Phi_d \right) \, , 
\end{eqnarray}%
\label{eq:vd-definitions}%
\end{subequations}%
and substitute the resulting expressions for $\dot\Phi_d$ and
$\dot\Phi_b$ into \eq{mass-conservation}. Solving for $\dot v_d$ one
obtains then after some straightforward algebra
\begin{equation}
  \dot v_d  = \frac{1}{2} \left( \zeta + \xi \varphi \right) \, .
  \label{eq:dot-vd}
\end{equation}
Here, $\varphi$ is the reduced average concentration defined in \eq{varphi}. It takes the
value $\varphi=1$ when $\phi=\Phi_b$, and smaller values for compositions
inside the miscibility gap. 

Assuming local equilibrium one can characterise the local bulk
concentration by the space dependent field $\varphi(x,t)$.
Its time evolution obeys a diffusion equation 
with a source strength of $2\, \dot v_d$
\citep{cates03PhilTrans}. According to the above
consideration this source term gives rise to a corresponding
growth of the equilibrium droplet volume fraction. 
From the point of view of the transport equations, the magnitude of
the source strength $\dot v_d$ appears therefore as the relevant
parameter characterising how strongly the mixture is driven away from
equilibrium. With this motivation we consider here temperature
protocols that correspond to fixed values of~$\dot v_d$.

In \eq{vd-definitions} it is understood that $\Phi_0$ and $\bar\Phi$
are functions of $T(t)$ due to their dependence of $\Phi_d$ and
$\Phi_b$, \ie on the borders of the two-phase region of the phase
diagram. In general these functions have a different temperature
dependence.  Hence, it is not clear a priory that $\dot v_d$ can be
fixed to a constant value by choosing an appropriate form of the
temperature ramp $T(t)$.  Indeed, we choose different temperature
protocols for the two phases\@{}---\@{}\ie for the M+H (and IBE+W) mixtures
we adopt different temperature ramps for methanol (IBE) droplets in
hexane (water) than for hexane (water) droplets in methanol (IBE).
The optimal protocol is found by rearranging \eq{dot-vd} to take the
form
\begin{equation}
   \dot v_d
   =
   \frac{1}{2 \Phi_{0}(T)}  
   \frac{\rmd\Phi_b}{\rmd t}  
   - \frac{v_d}{\Phi_0} \; \frac{\rmd \Phi_0}{\rmd t}
   \approx
   \frac{1}{2 \Phi_{0}(T)}  
   \frac{\rmd \Phi_b}{\rmd T}  
   \frac{\rmd T}{\rmd t} \, ,
\label{eq:xi}
\end{equation}
where the approximation in the final step is based on the fact that
the volume fraction of droplets $v_d$ is always small in our experiments.
According to \eq{xi} the ramp rate $\dot v_d$ for droplets in the
upper and lower layer of our samples is found by appropriately
assigning the indices $b$ and $d$ to the respective branches of the
phase diagram. Subsequently, the temperature protocol of the ramp is
obtained by integrating
\begin{equation}
\frac{\rmd T}{\rmd t} = 
   2\, \dot v_d
   \; \Phi_{0}(T)
   \; \left( \frac{\rmd \Phi_d}{\rmd T} \right)^{-1} \, .
\label{eq:calcT}
\end{equation}

In practice there is only a small difference between $\dot v_d$ and
$\xi$ since $\zeta \ll \xi$ for the phase diagrams under
consideration, and since $\varphi$ is always very close to one.
Hence, on the one hand, we distinguish between $\dot v_d$ and $\xi$
for the sake of calculating the temperature protocol.  This avoids
systematic errors in the numerical integration of \eq{calcT}.
On the other hand, for the further presentation of the data, we specify the ramp
rate in terms of $\xi$. 
This allows us to use terminology that is
consistent with the pertinent
literature \citep{cates03PhilTrans, vollmer99, auernhammer05JCP, vollmer07PRL, LappRohloffVollmerHof2012}.

\begin{figure}
\includegraphics[width=\textwidth]{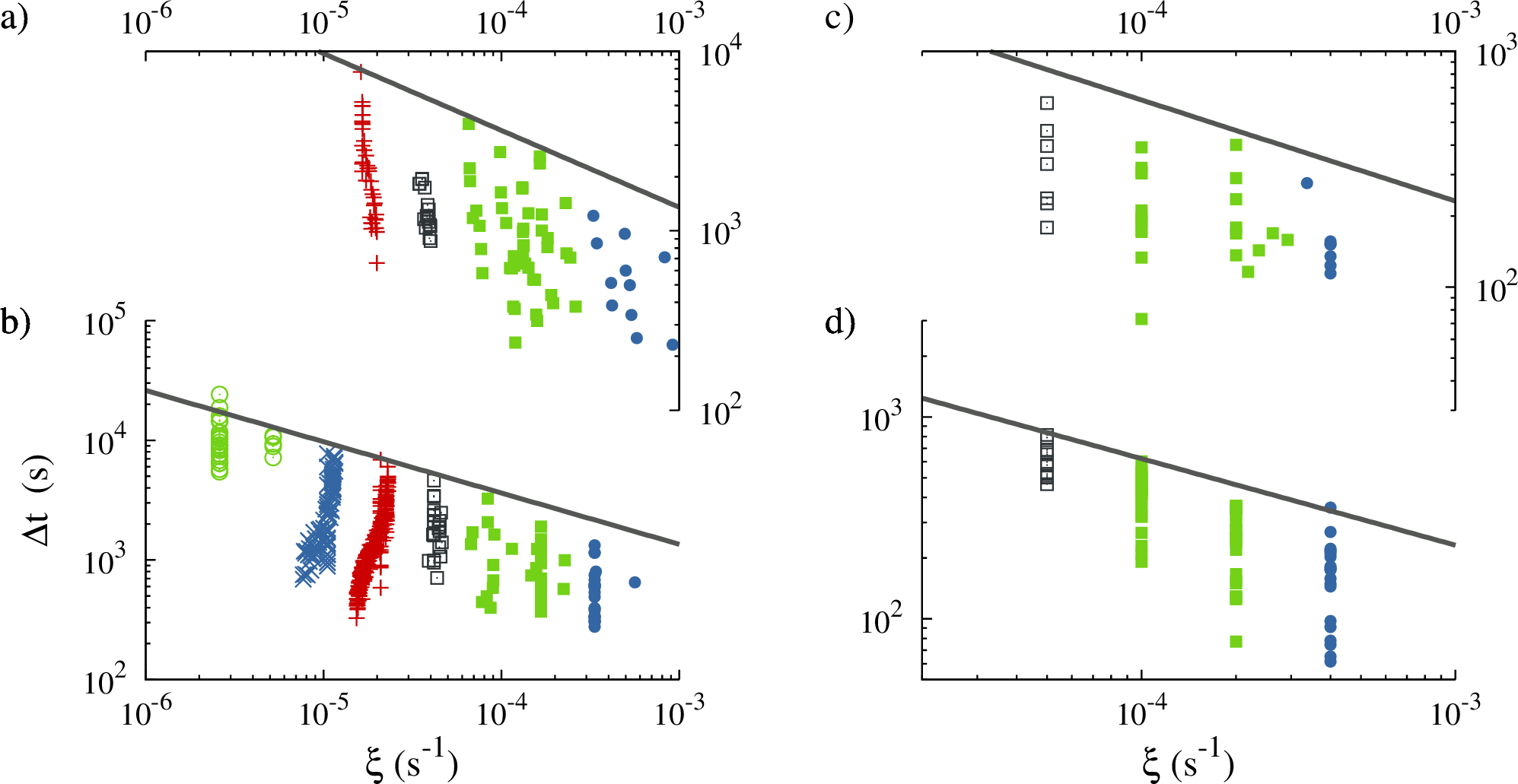}
\caption[]{\textbf{The oscillation period, $\Delta t$, plotted vs.~the ramp rate $\xi$.}
  The four panels show $\Delta t$ 
  for 
  a)  water-rich droplets in an IBE-rich continuous phase, 
  b)  IBE-rich droplets in a water-rich continuous phase, 
  c)  methanol-rich droplets in a hexane-rich continuous phase, and
  d)  hexane-rich droplets in a methanol-rich continuous phase,
  respectively. The different symbols denote measurements for
  different ranges of ramp rates, $\xi$. 
  The colours and symbols encode different heating rates $\xi$: 
  open green circle, $\xi < 6 \times 10^{-6}\,$s$^{-1}$; 
  blue cross, $10^{-6}\,$s$^{-1} < \xi < 1.3\times 10^{-5}\,$s$^{-1}$; 
  red plus, $1.3\times 10^{-5}\,$s$^{-1} < \xi < 3\times 10^{-5}\,$s$^{-1}$; 
  open black square, $3 \times 10^{-5}\,$s$^{-1} < \xi < 6\times 10^{-5}\,$s$^{-1}$; 
  green square, $6 \times 10^{-5}\,$s$^{-1} < \xi < 3\times 10^{-4}\,$s$^{-1}$; 
  and blue circle, $3 \times 10^{-4}\,$s$^{-1} < \xi$. 
  The grey lines are guides to the eye that indicate the slope of a
  power law, $\Delta t \sim \xi^{-3/7}$.
  \label{fig:orig-data}}
\end{figure}

\subsection{Experimental results for $\Delta t$}

\FIG{orig-data} compiles data of $\Delta t$ for a
vast range of heating rates $\xi$, and four different scenarios of
phase separation in a binary mixture: 
a) the emergence and sedimentation of water-rich droplets in an
isobutoxyethanol-rich phase;
b) the emergence and rising of isobutoxyethanol-rich droplets in a
water-rich phase;
c) the emergence and sedimentation of methanol-rich droplets in a
hexane-rich phase; and
d) the emergence and rising of hexane-rich droplets in a
methanol-rich phase.

Different data points for a given ramp rate are due to the drift of
$\Delta t$ when pertinent material constants
change upon moving further away from the critical point.
In \App{MatConst} we provide the temperature dependence of the
material constants, which in turn translates to a time dependence when inverting the 
protocol $T(t)$ of the temperature ramp. 
For all data the height of the layer was $h\approx 1\,{\rm cm}$.
Measurements for samples with varying heights between $h=0.25\,{\rm cm}$ 
and $5.5\,{\rm cm}$ for the lower layer showed that $\Delta t$
is hardly affected by $h$.
The data points for the IBE+W mixture (left) are obtained by
particle tracking (\cf \citealt{LappRohloffVollmerHof2012} for experimental details), 
and those for M+H (right) refer to subsequent minima of turbidity
measurements as shown in \Fig{phasdias}. We verified that both methods
provide the same results. However, the data obtained from droplet
tracking tend to be more accurate.

In the following section we establish a model for the droplet growth
and sedimentation that provides a quantitative description of
$\Delta t$ for all data presented in \Fig{orig-data}.

\section{Theory} 
\label{sec:theory}

As a first step to model $\Delta t$ we consider the reasons why the 
turbidity --- and hence the precipitation rate --- 
in our experiment is not steady:
the turbidity of a transparent fluid mixture increases when a considerable
number of droplets have grown to a size comparable to (and eventually
larger than) the wavelength of light.  This manifests as a change of
colour in the lower part of the cell when the system progresses from
the snapshots shown in \Fig{spacetime}.b)--c).
Conversely, the fluid becomes clearer again when vast amounts of small
droplets are collected during the sedimentation of the largest droplets 
(transition from \Fig{spacetime}.d)--e)). 
Repetition of the cycle of nucleation, growth of droplets,
and resetting the system by sedimentation gives rise to episodic
precipitation, as shown in the space-time plot, \Fig{spacetime}.g).
In the following the salient features of this dynamics are modelled.

\subsection{Evolution of the radius of the largest droplets}

We start with general considerations motivating the setup of the model.

1. Spatial degrees of freedom need not be considered to describe
  the evolution of the largest droplets. For the nonlinear reactions
terms characterising phase separation the convective mixing
efficiently eliminates spatial inhomogeneities of the droplet size
distribution \citep{BenczikVollmer2010,BenczikVollmer2012}.  Indeed,
based on visual inspection of the accompanying movies, we estimate the
mixing time scale to be of the order of seconds. It is about three
orders of magnitude smaller than the period $\Delta t$.

2. It is sufficient to consider the characteristic size of the largest
droplets rather than the full droplet size distribution.  For
diffusively growing droplets the size distribution is sharply bounded
towards large droplets.  Consequently, the largest droplets in the
system have a well-defined size and there are only few of these
droplets
\citep{Slezov2009,ClarkKumarOwenChan2011,VollmerPapkeRohloff2014}.
When buoyancy starts to effect their motion these large droplets
collect smaller droplets, grow rapidly, and eventually clear the
system from droplets by precipitation \citep{KostinskiShaw2005}.

\begin{figure}
  \[
  \includegraphics[width=0.5\textwidth]{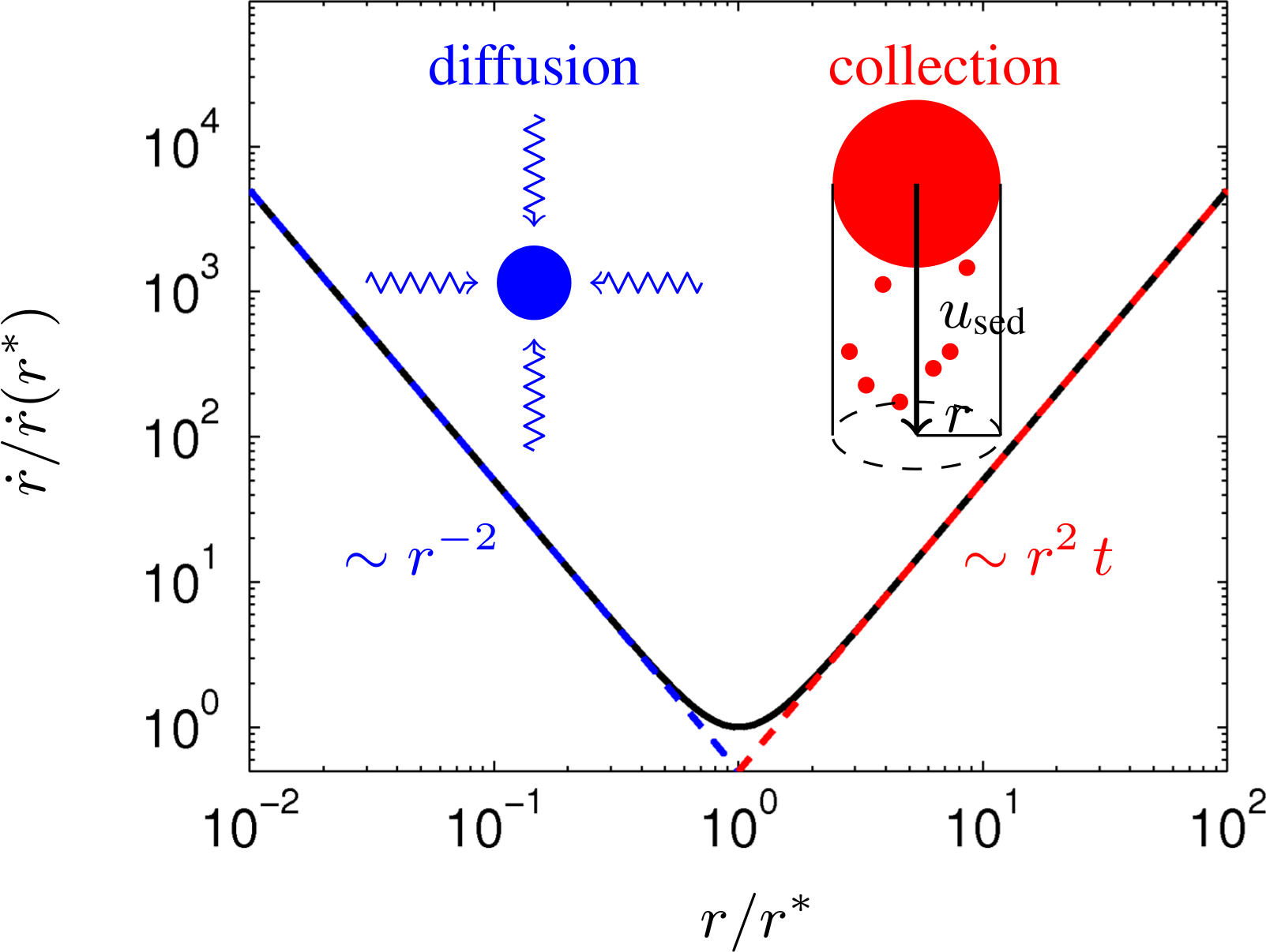}
  \]
  \caption[]{\textbf{Bottleneck in droplet growth.} 
    As a function of droplet size $r$ the growth speed $\dot r$ 
    of droplets shows a sharp minimum at a size $r^{\ast}$. Smaller
    droplets grow by diffusion --- the growth of larger droplets is speeded up by 
    sedimentation, that promotes 
    the collection of small droplets (see insets).}
  \label{fig:bottleneck}
\end{figure}

3. While many different processes contribute to the droplet growth, it
suffices to consider only droplet growth by diffusive accretion that
relaxes supersaturation, and the collection of small droplets by
sedimenting large ones in order to achieve a quantitative description
of $\Delta t$.  The processes are illustrated in \Fig{bottleneck}, and
we will now discuss them in turn.

\subsubsection{Growth by diffusive accretion that relaxes supersaturation}

The dynamics of large droplets crossing the meniscus
\citep{AartsDullensLekkerkerker2005} and droplet nucleation
\citep{BinderStauffer1976,FarjounNeu2011} provide microscopic droplets
in the fluid.  Subsequently, the supersaturation in the bulk relaxes
by diffusion of the minority component onto the droplets. The
diffusive accretion of material on the droplets relaxes
supersaturation and induces droplet growth.

In the experiments the temperature ramp 
is adjusted in such a way that the volume fraction of droplets grows linearly in time 
with a speed $\xi$.  
For these growth conditions it was demonstrated in
\citet{Sugimoto1992,TokuyamaEnomoto1993,ClarkKumarOwenChan2011,VollmerPapkeRohloff2014} that the number density of droplets is
preserved.
Droplets of a characteristic radius $r$ and number
density $\dichte$ occupy a volume fraction $\dichte \, 4\pi r^3/3$.
When the droplet volume fraction increases with speed $\xi$ and the number density $\dichte$
is conserved, diffusive growth provides a temporal change of the droplet radius
\begin{equation}
  \frac{\rmd}{\rmd t} \frac{4\pi \, n \, r^3}{3} = \xi 
  \quad \Rightarrow \quad
  \dot r = \frac{\xi}{4\pi \dichte} \; \frac{1}{r^2} \; .
  \label{eq:accretion-rate}
\end{equation}
Alternatively, this growth law can be obtained as
large $k$ approximation of 
the diffusive growth law \citep{ClarkKumarOwenChan2011,VollmerPapkeRohloff2014}
\begin{equation}
   \dot a = \frac{\sigma D}{a^2} \left( k \, \frac{a}{\Rav} - 1 \right) \, ,
   \quad
   k = 1 + \frac{ \xi }{4\pi \sigma D n}
\label{eq:r-dot-frontier}
\end{equation}
describing the growth of a droplet of radius $a$ in an assembly
of droplets with distribution $P(a)$ and mean droplet radius 
$\Rav = \int \rmd a \, a \, P(a)$.  
In \eq{r-dot-frontier} $D$ is the diffusion coefficient for accretion
of material on the droplets, and
\begin{equation}
  \sigma = \frac{2 \gamma V_m^2 C_{\infty}}{R T} 
  \label{eq:sigma}
\end{equation}
is the Kelvin length \citep{Lif+81,Bray1994}, that depends on the
interfacial tension $\gamma$, the molar volume $V_m$, and the
equilibrium composition of the droplet phase $C_{\infty}$ in units of
mol/m$^3$.  (Specific values of the material constants are provided in \App{MatConst}.)  
It was shown in \citet{VollmerPapkeRohloff2014} that $k$ takes values of the
order to $10^6$ under the conditions considered here, 
and that $a \simeq \Rav$ in the late stages of competitive droplet growth
at large $k$. Hence, \eq{r-dot-frontier} reduces to \eq{accretion-rate}.

\subsubsection{Growth by collection of smaller droplets}

When the droplets become sufficiently large, they 
drift under the influence of buoyancy forces. According to Stokes'
formula the velocity of a slowly settling 
droplet is \citep{TaylorAcrivos1964,guyonBook}
\begin{equation}
  u=\kappa r^2  
  \qquad \textrm{with} \quad 
  \kappa = \frac{2}{9} \; \frac{ g \, \Delta \rho}{ \visc_b} \; 
  \frac{\visc_d + \visc_b}{\visc_d + \frac{2}{3}\visc_b}  \: ,
  \label{eq:u}
\end{equation}
where $g$ is the gravitational acceleration, $\Delta \rho$ the density
contrast, $\visc_b$ is the dynamic viscosity of the bulk phase, and
$\visc_d$ is the viscosity of the material in the droplets.  When the
Stokes velocity of the the largest droplets in the system becomes
noticeable they collect smaller droplets in their path.  Hence, the
volume of a large droplet grows like $4\pi r^2 \, \dot r =
\collEfficiency \pi r^2 u \xi t$, where $\collEfficiency $ is the
collection efficiency for large droplets coalescing with smaller ones,
and $\xi t$ is the volume fraction of the smaller droplets.  (Observe
that $r$ refers to the radius of the largest droplets in the
system\@{}---\@{}a minute minority of droplets that accounts for
only a small part of the droplet volume fraction.)
Accordingly, we find the collisional growth rate 
\begin{equation}
  \dot r =\frac{\collEfficiency\kappa\xi t}{4}r^2 \;.
  \label{eq:sweep-rate}
\end{equation}

\subsubsection{The bottleneck of droplet growth}

The diffusive growth mechanism, \eq{accretion-rate}, works very well
for small droplets due to the factor $r^{-2}$, and it becomes less and
less efficient when $r$ grows.  In contrast, growth by collecting
small droplets, \eq{sweep-rate}, does not contribute to the growth as
long as all droplets are small, while it leads to runaway growth of
the large droplets when their motion is affected by buoyancy.  Hence,
we assert that the sum of the diffusive growth, \eq{accretion-rate},
and the contribution accounting for the collection of smaller
droplets, \eq{sweep-rate},
\begin{equation}
  \dot r=\frac{\xi}{4\pi \dichte}\frac{1}{r^2} + \frac{\collEfficiency\kappa\xi t}{4}r^2 \, ,
  \label{eq:r-dot}
\end{equation}
faithfully describes the growth of the largest droplets in the system.
The growth law, \Eq{r-dot}, shows a bottleneck of growth at the bottleneck radius,
$r^\ast$, where the droplet growth speed, $\dot r$, takes its smallest
value, $\dot r(r^\ast,t^\ast) = (2/3) \left( r^\ast \right)^{-2} $
(see \Fig{bottleneck}), 
\begin{subequations}
\begin{equation}
  r^\ast = r(t^\ast) \simeq \left( \pi \dichte \collEfficiency \kappa \, t^\ast \right)^{-1/4} \, .
  \label{eq:r-ast}
\end{equation}
The bottleneck is approached at the time $t^\ast$ required for
droplets to grow from zero radius to the radius $r^\ast$.  Integrating
\eq{accretion-rate} from $r=0$ to $r=r^\ast$ yields $4\pi \dichte
{r^{\ast}}^3 / 3 = \xi t^{\ast}$.  Together with \Eq{r-ast} this
equation provides the following expressions for the bottleneck time
$t^{\ast}$ and the bottleneck radius $r^\ast$,
\begin{equation}
  t^\ast = \left(\frac{2^8 \, \pi}{3^4} \; \frac{\dichte}{\collEfficiency^3\kappa^3 \xi^4}\right)^{1/7}  \, ,
  \qquad 
  r^\ast = \left( \frac{3}{4\pi^2}\frac{\xi}{\collEfficiency \kappa \dichte^2}\right)^{1/7} \, .
\label{eq:t-ast}
\end{equation}
\label{eq:dim-less-units}%
\end{subequations}%

Henceforth, we measure time in units of  $t^\ast$, 
droplet radii in units of  $r^\ast$, 
and, for conciseness of the notation, we denote the resulting
dimensionless units still as $(r,t)$.  
In terms of these dimensionless
variables \eq{r-dot} takes the form
\begin{equation}
  \dot r = \frac{1}{3 \, r^2} + \frac{t \, r^2}{3} \, ,
\label{eq:rDotDL}
\end{equation}
such that the growth velocity $\dot r(r,t)$ takes its minimum at $(r,t) =
(1,1)$.

\subsection{Calculating the period $\Delta t$}
\label{sec:period}

\begin{figure}
\[ \includegraphics{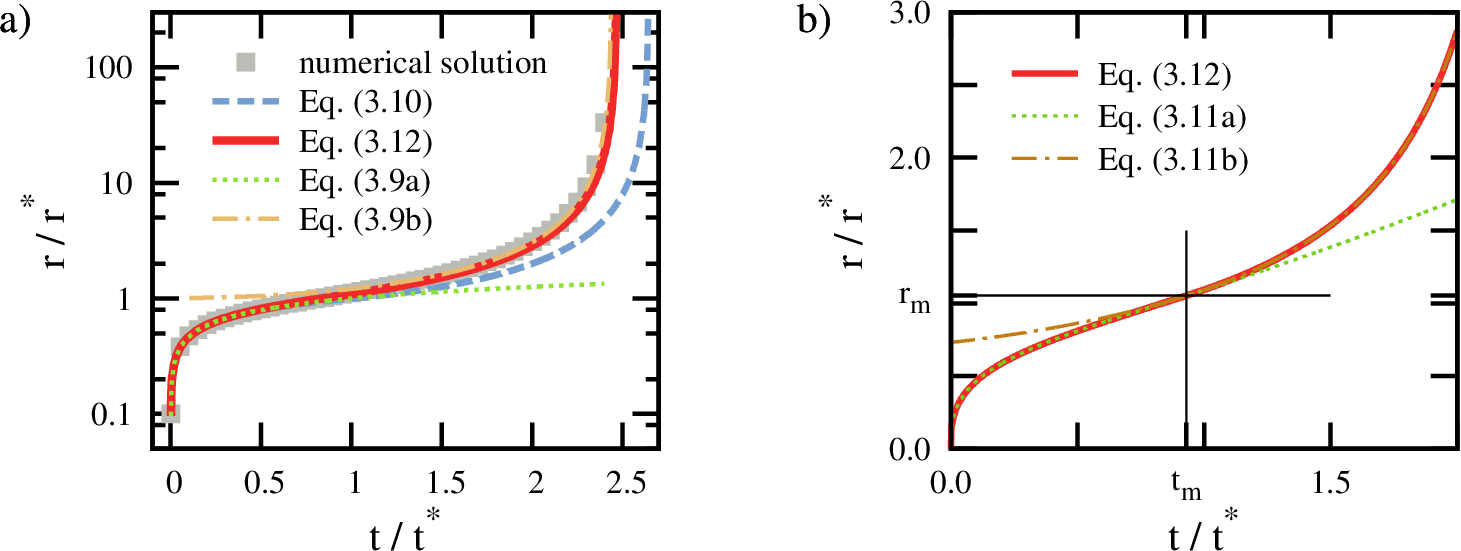} \]
\caption[]{\textbf{Finite-time divergence of droplet growth.} 
  a) Comparison of the numerical solution of \eq{rDotDL} (grey squares)
  to the analytical approximations, 
  \eq{approx} (dashed blue line) and 
  \eq{refined} (solid red line), 
  respectively. 
  The green dotted line shows the leading
  order approximation for small droplets, \eq{rS1}, 
  and the brown dash-dotted lines the description of the divergence of the 
  size of large droplets, \eq{rL1}, evaluated for $\Delta t = 2.44$.
  The analytical description of the droplet growth, 
  (solid red line), is also shown in panel 
  b) which shows how \eq{refined} is obtained by matching the expressions \eq{rS} and
  \eq{rL} for small and large droplets, respectively.
  \label{fig:drop_size}
}
\end{figure}

As long as buoyancy does not yet affect the motion of the largest
droplets in the system, the droplets grow diffusively by collecting
supersaturation. In leading order for small droplets one can then
neglect the growth contribution $t r^2 /3$ in \Eq{rDotDL}. For an
initial droplet size $r(t=0)=0$ this entails
\begin{subequations}
\begin{equation}
  \label{eq:rS1}
  \dot r \simeq \frac{1}{3\,r^2}  \quad \Rightarrow  \quad r_S(t) \simeq t^{1/3} \, ,
\end{equation}
where the index $S$ in $r_S(t)$ stresses that the approximation applies as long as droplets are small, $r_S \lesssim 1$.
As shown by the dotted line in \Fig{drop_size}.a) this approximation
provides a good estimate for values $t < 1/2$. 

Similarly, for large droplets the contribution $(3 r)^{-2}$ to the
growth is sub-dominant in \Eq{rDotDL} such that in leading order
\begin{equation}
  \dot r \simeq \frac{r^2 \, t}{3}  
  \quad \Rightarrow  \quad 
  r_L(t) \simeq \frac{6}{\Delta t^2 - t^2} \, .
  \label{eq:rL1}
\end{equation}%
\end{subequations}%
Here, the index $L$ in $r_L(t)$ indicates that this solution applies
when the droplets are large, $r_L \gtrsim 1$.  The growth law,
\eq{rL1}, features a finite-time singularity when $t$ approaches
$\Delta t$. At the latest at this late time, the large droplets will
rapidly fall out of the measurement window, such that the system is
reset to its initial state $r\simeq 0$.  On the one hand, the
dash-dotted line in \Fig{drop_size}.a) shows that \eq{rL1} provides a
very good description of the numerical data for $t \gtrsim t^\star$
for the choice $\Delta t = 2.44$.  On the other hand, the expression
\eq{rL1} can not be matched continuously to \eq{rS1} because for
$\Delta t = 2.44$ the latter expression produces smaller values for
$r(t)$ for all $t$.
Rather, a continuous and differentiable interpolation from \eq{rS1} to
\eq{rL1} requires to choose
\begin{equation}
  r(t) \simeq \left\{
    \begin{array}{l@{\quad\textrm{for}\quad}l}
      t^{1/3}               & t \leq 1 \, , \\[1mm]
      6 \; ( 7 - t^2 )^{-1} & t \geq 1 \, .
    \end{array}\right .
  \label{eq:approx}
\end{equation}
The resulting first order estimate for $r(t)$ is shown by the dashed
blue line in \Fig{drop_size}.a). It diverges at $\Delta t = \sqrt{7}
\simeq 2.646$, thus overestimating the time $\Delta t$ required to
reach the finite-time singularity observed in the numerical data by
about $8$\%.

A more accurate description of the numerical solution of \eq{rDotDL}
is obtained by taking into account the leading order corrections of
\eq{rS1} and \eq{rL1}.
A refined estimate for the droplet growth is obtained by using
$r_S(t)$ to approximate the sub-leading contribution 
to the growth of $r^3$ by $t r^4 \simeq t^{7/3}$. The resulting
solution of \eq{rDotDL} becomes
\begin{subequations}
\begin{equation}
  \frac{\rmd}{\rmd t} r^3 \simeq 1 + t^{7/3}  
  \quad \Rightarrow  \quad 
  r_s(t) \simeq \left( t + \frac{3}{10} \; t^{10/3} \right)^{1/3} \, .
  \label{eq:rS}
\end{equation}
This expression provides an excellent fit to the numerical data for $t
\lesssim t^\star$, as shown by the the dotted green line in 
of \Fig{drop_size}.b).

\begin{figure}
  \[  \includegraphics{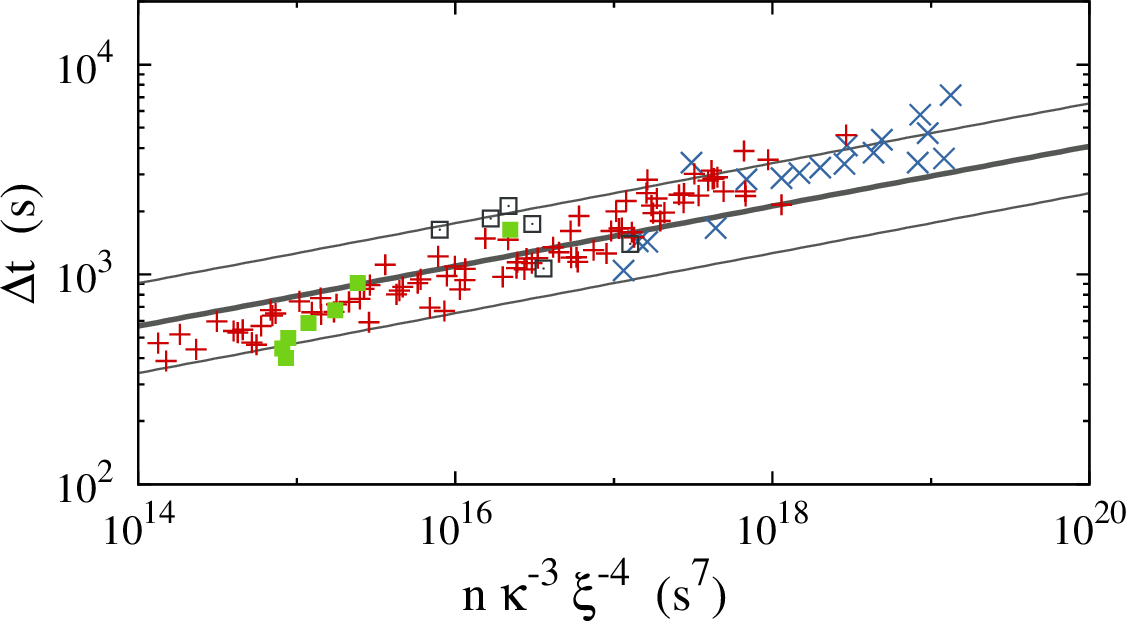} \]
  \caption[]{\textbf{Dependence of $\Delta t$ on $\dichte$, $\kappa$, and $\xi$.}
    The period of episodic precipitation in IBE+W
    mixtures of various compositions collapse on a curve when plotting 
    $\Delta t$ as a function of $\dichte \kappa^{-3} \xi^{-4}$.  
    Here, the respective heating rates in units of s$^{-1}$ are
    specified in the caption of \Fig{orig-data} and the measurement of the number
    densities, $n$, have been reported in \citet{LappRohloffVollmerHof2012}. 
    When collecting the numerical prefactors and accounting for the
    temperature dependence of $\kappa$, the collection efficiency,
    $\collEfficiency$, remains as the only free parameter of the
    prediction \Eq{DeltaT37}. Varying $\collEfficiency$ induces a
    vertical displacement of the line. Here, $\collEfficiency$ was set to the value
    $\collEfficiency=0.3$ for the thick solid line. The thinner
    lines above and below correspond to the choices $\collEfficiency =
    0.1$ and $\collEfficiency = 1$, respectively.}
  \label{fig:DeltaT_n0-scaling}
\end{figure}

For the large droplets a more accurate prediction is obtained by using
$r_L(t)$ to approximate the $(3r)^{-2}$ term in \eq{rDotDL},
\begin{eqnarray}
  -\frac{\rmd}{\rmd t} r^{-1}  
  & = & \frac{t}{3} 
       +  \frac{1}{3} \; r^{-4}
  \simeq  \frac{t}{3} 
       +  \frac{1}{3} \; \left( \frac{\Delta t^2 - t^2}{6} \right)^4
  \nonumber \\[2mm]
  \Rightarrow  \; 
  r_l(t, \Delta t) 
  & \simeq &
  \left[ \frac{\Delta t^2 - t^2}{6} \right.
  \label{eq:rL}
    \\ &&  \left.
    + \frac{(\Delta t-t)^5}{3\cdot 6^4} \; 
    \left(\frac{t^4}{9} 
      + \frac{5 \, t^3 \, \Delta t}{9} 
      + \frac{23 \, t^2 \, \Delta t^2}{21} 
      + \frac{65 \, t \, \Delta t^3}{63} 
      + \frac{128 \, \Delta t^4}{315} \right) 
    \right]^{-1} \, .
    \nonumber
\end{eqnarray}%
\end{subequations}
When evaluated at $\Delta t = 2.467$  the expressions \Eq{rS} and \Eq{rL}
match continuously and differentiable 
at the point $(t_m,r_m) = (0.9304,1.0526)$,
[\Fig{drop_size}.b)]
\begin{equation}
  \label{eq:refined}
  r(t) \simeq \left\{\begin{array}{l@{\quad\textrm{for}\quad}l}
    r_s(t)                      & t \leq t_m = 0.9304 \, , \\[1mm]
    r_l(t, \Delta t = 2.4667)   & t \geq t_m = 0.9304 \, .
\end{array}\right .
\end{equation}
The thick solid red lines in both panels of \Fig{drop_size} show the
expression \Eq{refined} over the full $t$-range. It provides an excellent
description of the numerical solution of \eq{rDotDL} that is shown by grey squares.
In particular, the position of the predicted
finite-time singularity, $\Delta t = 2.467$, is only off by one
percent from the numerically obtained value, $\Delta t = 2.44$.

In conclusion, the parameter dependence of the time scale, $\Delta t$,
for the growth from vanishingly small to very large droplets is
provided by the time, $t^\ast$, required to grow to the bottleneck
size, $r^\ast$.  Based on \eq{t-ast} and the fit of $\Delta t$ in
\eq{rL1} to match the asymptotics of the numerical data shown in
\Fig{drop_size}, we find
\begin{equation}
  \Delta t 
 \simeq 2.44 \, t^\ast 
 \simeq 3.39 \; \left( \frac{n}{\collEfficiency^3 \kappa^3 \xi^4}\right)^{1/7}
  \ .
\label{eq:DeltaT37}
\end{equation}
A first hint that this prediction might be faithful is obtained by
observing that the parts of a period where we observe high and low
turbidity in \Fig{phasdias} are of comparable extent.  This is
consistent with the theoretical prediction that the singularity arises
at $\Delta t = 2.44 \, t^\ast$.
A more thorough test is presented in \Fig{DeltaT_n0-scaling} where we
plot $\Delta t$ as function of $n \kappa^{-3} \xi^{-4}$. 
The plot
is based
on 
data of \citet{LappRohloffVollmerHof2012} where the time evolution of
the droplet density was followed by particle tracking such that both,
$\Delta t$ and \dichte, are known from the experiment.  The data
determining the temperature dependence of $\kappa$ is provided in
\App{MatConst}.  Hence, the collection efficiency, $\collEfficiency$,
remains as the only free parameter of the prediction,
\eq{DeltaT37}. It induces a vertical displacement of the prediction on
the logarithmic scale in \Fig{DeltaT_n0-scaling}.
The theoretical curves displayed in \Fig{DeltaT_n0-scaling} show the
prediction \eq{DeltaT37} for the constant values, $\collEfficiency =
0.1$, $0.3$, and $1$, respectively.  These values correspond to the
middle and the respective most extreme values observed for other
systems \citep{BeardOchs1993}, where the collection efficiency
$\collEfficiency$ was reported to take values in the range $0.1 \leq
\collEfficiency \leq 1$.  All data points lie in the narrow band
around the prediction, well within 
the uncertainty of $\collEfficiency$.  There only is a slight
systematic mismatch of the slope. We attribute this trend to a weak
temperature dependence of $\collEfficiency$.  The mismatch arises from
a correlation of the temperature dependence of $\kappa$ and $\alpha$,
and the corresponding dependence of $\collEfficiency$.

\subsection{Accounting for different droplet densities}

\begin{figure}
  \includegraphics[width=\textwidth]{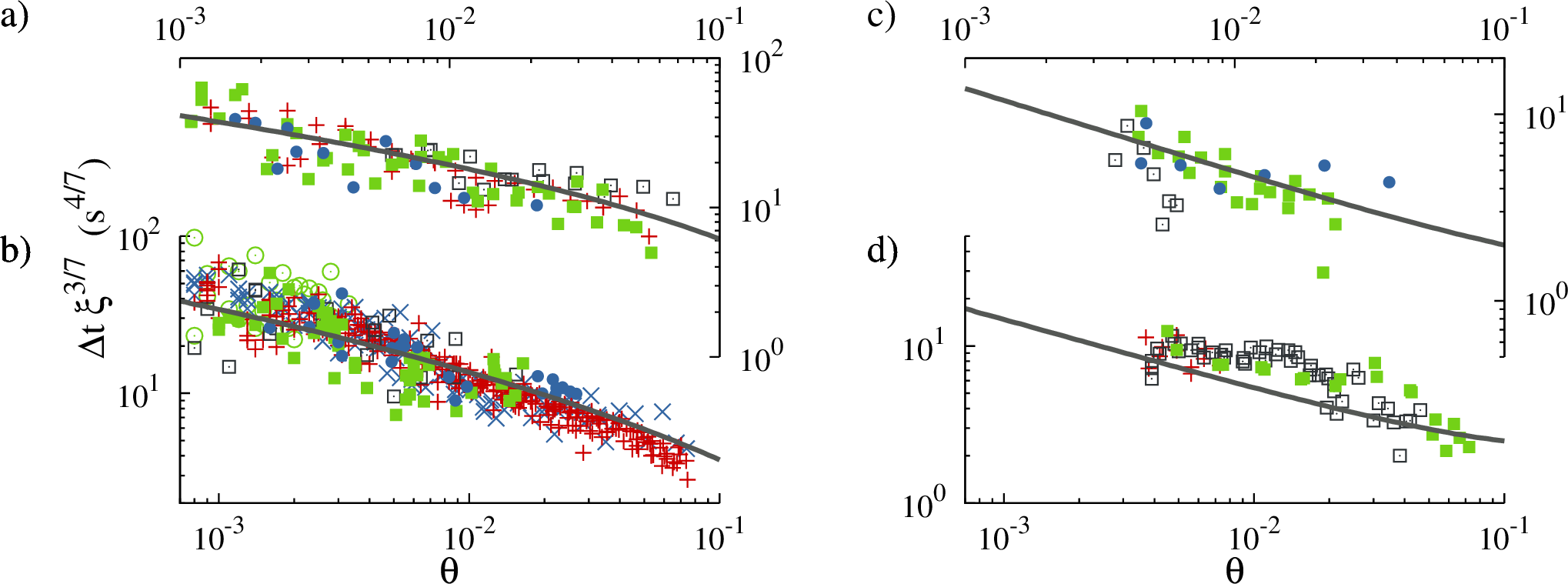}
  \caption[]{\textbf{Temperature dependence of $\Delta t \: \xi^{3/7}$ for IBE+W and M+H mixtures.} 
    Data points are shown for the upper (top) and the lower layer
    (bottom) of mixtures of IBE+W (left) and M+H (right),
    respectively.  We use the same symbols and colours as in
    \Fig{orig-data}, and show the theoretical prediction,
    \eq{Delta_t}, by solid lines.  The same value of $\alpha$ is found
    for the upper and the lower layer of the mixtures, $\alpha=0.71$
    for IBE+W (left), and $\alpha = 0.9$ for M+H (right).  }
  \label{fig:master}
\end{figure}

Typically the droplet density,
$\dichte$, is not easily accessible.  It is therefore desirable to provide
an estimate for $\dichte$ in order to arrive at a widely applicable theory
for the period, $\Delta t$.
In the context of our experiments this can be achieved by observing that
the expression $k$, defined in \eq{r-dot-frontier}, is preserved
during the periods of diffusive droplet growth
\citep{KleinMoisar1963,Sugimoto1992,ClarkKumarOwenChan2011},
and that $k \gg 1$ for the present experiments \citep{VollmerPapkeRohloff2014}.
Hence, the number density \dichte\ is
proportional to $\xi / ( \sigma D )$.
Combining this proportionality with \eq{DeltaT37} we obtain
\begin{eqnarray}
  \Delta t  &=&  \alpha   \left({D\sigma\kappa^3}\right)^{-1/7}\xi^{-3/7} \, .
  \label{eq:Delta_t}
\end{eqnarray}%
The factor $\alpha$ comprises numerical prefactors and the dependence
of $\Delta t$ on quantities that are not accessible in many
circumstances: the collection efficiency, $\collEfficiency$, and the
parameter $k$ characterising the diffusive growth.
The coefficients $D$, $\sigma$ and $\kappa$ in \eq{Delta_t} are functions of material
constants. They show a strong temperature dependence that arises
from the vanishing of the interfacial tension and the mass density
contrast at the critical temperature, $T_c$, of the phase transition.
This, in turn, entails the vanishing of $\sigma$ and $\kappa$ which
are proportional to the interfacial tension and the mass density contrast,
respectively (\cf\App{MatConst}).
Hence, \eq{Delta_t} suggests that $\Delta t \:
\xi^{3/7}$ should be a function of
the reduced temperature $\theta = |T - T_c|/T_c$.
This proposition is corroborated in \Fig{master}. It shows a
remarkable data collapse for all data compiled in \Fig{orig-data} when
plotting $\Delta t \, \xi^{3/7}$ as function of $\theta$.  Moreover,
the resulting temperature dependence is faithfully described by the
master curves, \eq{Delta_t}.  The dimensionless prefactor $\alpha$ is
the only free parameter in this description.  This parameter takes
values very close to unity that only depend on the selected mixture:
$\alpha=0.71$ for IBE+W (left panels of \Fig{master}), and
$\alpha=0.9$ for M+H (right panels of \Fig{master}).

\section{Discussion}
\label{sec:discussion}

In the present section we interpret the modeling of the data for the
binary mixtures with particular emphasis on the quality of the data
collapses shown in \Figs{DeltaT_n0-scaling} and \fig{master}. What are
the underlying assumptions?
What would one expect for other systems?

\subsection{Values of $\alpha$ for binary mixtures}

The solid lines in \Fig{master} faithfully provide the $\theta$
dependence of $\Delta t$ even though \Eq{Delta_t} only accounts for the
temperature dependence of the material constants, and disregards the
temperature dependence of $\collEfficiency$ and $k$, that should be
present according to our discussion of \Fig{DeltaT_n0-scaling} in
\Sect{period}.
Consequently,
the dimensionless prefactor $\alpha$ is the only free parameter in
\eq{Delta_t}. Comparing \eq{DeltaT37}, \eq{r-dot-frontier} and
\eq{Delta_t} one finds,
\begin{equation}
  \alpha 
  \simeq  2.44 \; \left( \frac{2^6}{3^4} \right)^{1/7} \;  (k-1)^{-1/7} \, \collEfficiency^{-3/7}  \, ,
  \label{eq:def-alpha}
\end{equation}
where $2.44$ is the ratio of $\Delta t$ and the bottleneck time scale
$t^{\ast}$ found by the fit of \Eq{rL1} in \Fig{drop_size}.  Typical
values of $\collEfficiency$ are $0.1 \lesssim \collEfficiency \lesssim
1$ \citep{BeardOchs1993}, and for the IBE+W system
\citet{VollmerPapkeRohloff2014} reported $10^5 \lesssim k \lesssim
10^7$.  For $k=1.7 \times 10^5$ and $\collEfficiency = 0.3$ one indeed
finds the value $\alpha = 0.71$ adopted in \Fig{master}.
There is only a weak variability of $\alpha$ in spite of the
substantial range of values taken by $\collEfficiency$ and $k$: the
$(1/7)^{\textrm{th}}$ and $(3/7)^{\textrm{th}}$ power in
\eq{def-alpha} strongly suppress these dependences.

\subsection{Temperature dependence of $\Delta t$}

In contrast to a suggestion in the literature \citep{Wilkinson2014} we
are reluctant to attribute the $\theta$ dependence of $\Delta t \: \xi^{3/7}$ to
the critical scaling of the material constants 
entering \eq{Delta_t}, \ie the dependence on 
$D$, $\sigma = 2 \gamma V_m^2 C_{\infty}/(R T)$ and $\kappa = 2g\,\Delta\rho/(9\mu)$. 
The reason is fourfold:
\begin{itemize}
\item[(i) ] the values of $\theta$ in our experiments clearly lie
  outside the critical range. This is documented in the \App{MatConst}
  where we report a much more involved $\theta$ dependence of the
  material constants than the power-law singularities describing the
  scaling for small reduced temperatures $\theta$;
\item[(ii) ] in addition to $D$, $\sigma$ and $\kappa$ also the
  collection efficiency $\collEfficiency$ shows a noticeable temperature
  dependence, as observed in \Fig{DeltaT_n0-scaling};
\item[(iii) ] the parameter $k$ entering the definition,
  \eq{def-alpha}, of $\alpha$ has a noticeable temperature dependence
  \citep{VollmerPapkeRohloff2014};
\item[(iv) ] for the mixtures under consideration the dependence of
  $\collEfficiency$ and $k$ cancels partially.
\end{itemize}
Consequently, the close correspondence of the $\theta$ dependence of
the prediction \eq{Delta_t}, and the one obtained by considering
$\alpha$ to be a constant and $D$, $\sigma$ and $\kappa$ to vary
according to the power laws valid very close to critical point might
very well be a coincidence.
A proper discussion of the temperature dependence of $\Delta t$ should
first address the intriguing observation that $k$ takes surprisingly
large values in the present experiments, and that the observed values
vary so little that their dependence need not be considered to obtain
a good estimate of the oscillation period, \Fig{master}.

\subsection{Bottleneck radius}

\begin{figure}
  \includegraphics[width=\textwidth]{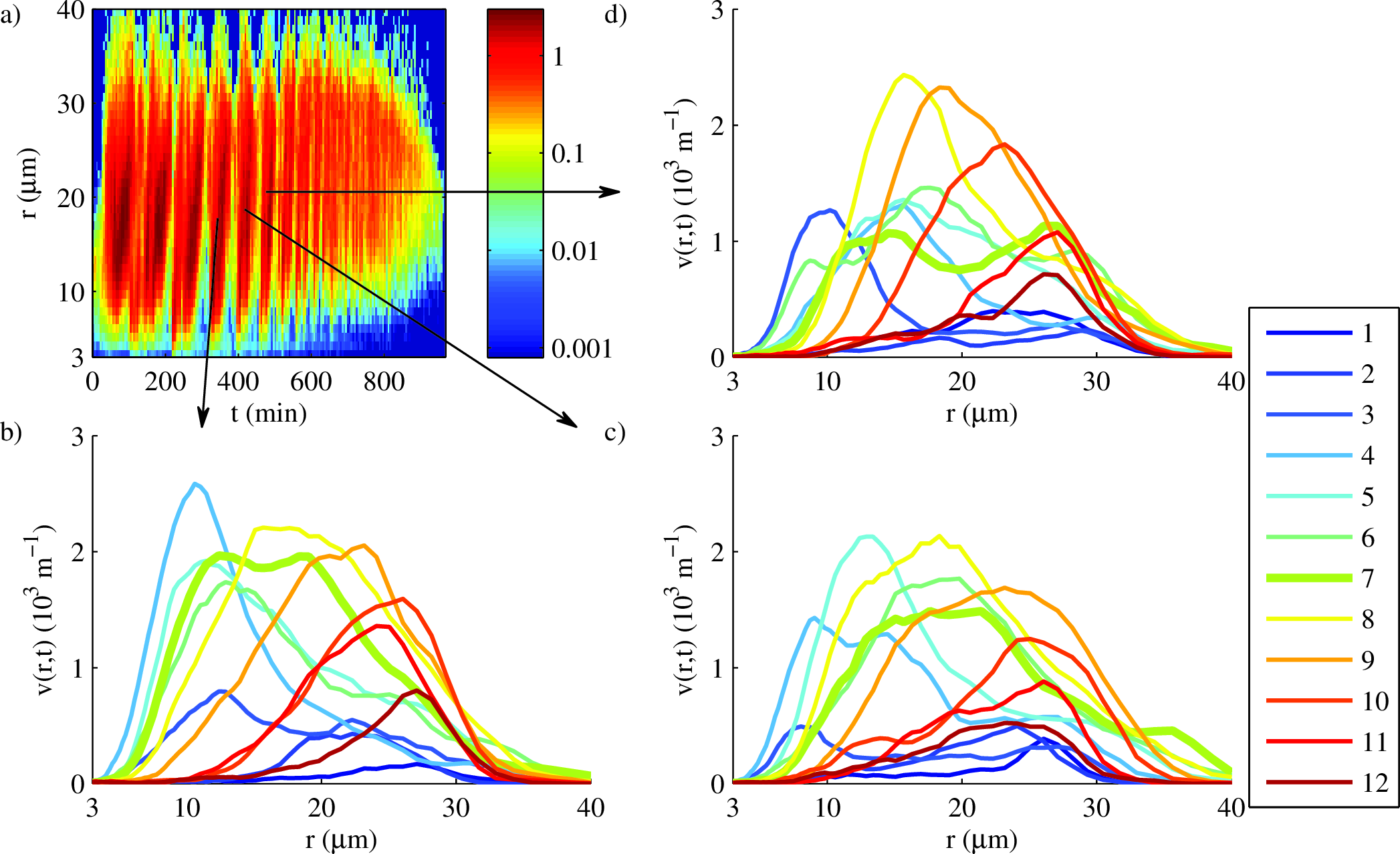}
  \caption[]{\textbf{Evolution of the droplet size distribution} for
    the lower layer of the IBE+W mixture subjected to a ramp rate of
    $\xi = 1.05 \times 10^5$s$^{-1}$.  a) A radius-time plot of the
    distribution of the droplet volume fraction $v(r,t)$ clearly
    captures the oscillations in time.  b--d) explicitly provide the
    radial distribution of the volume fraction for the fourth, fifth
    and sixth oscillation.  To suppress fluctuations the distributions
    are determined as temporal averages over one of twelve time
    intervals of equal length in each oscillations. Within each period
    the distributions at different times are labelled by a colour
    coding ranging from blue to red, as specified in the legend.
    The thicker green lines, number 7, correspond to the time
      where the bottleneck is crossed.
    \label{fig:sizedistribution}}
\end{figure}

\FIG{sizedistribution} shows the time evolution of the distribution of
the droplet volume fraction $v(r,t)$ of droplets of radius $r$. Panel
a) provides an overview in terms of a radius vs.~time plot where
$v(r,t)$ is indicated by false colour.  Each of the panels b)--d)
shows twelve curves that describe the evolution of the distribution
during one oscillation.  In the beginning of each period there is a
pronounced peak for small radii (blue lines).  The maximum of the
distribution shifts to larger radii as the distribution evolves, it
develops a shoulder (curve 3--5), becomes bimodal (thick green curve
7), and then the number of large droplets rapidly decays (curves
8--12).  We attribute the decay to precipitation.  The arising of the
shoulder reflects the broadening of the distributions when the largest
droplets have crossed the bottleneck
\citep{BeardOchs1993,KostinskiShaw2005}.
From this perspective the minimum arising in the bimodal droplet spectra 
should amount to the bottleneck radius, $r^\ast$. 
For the data of measurements in the lower layer of IBE+W, that are
shown in \Fig{sizedistribution}, the bottleneck radius is thus found
to lie in the range $r^{\ast} \simeq 15 \dots 20\,\mu$m [\cf the thick
green curves, number 7, in \Fig{sizedistribution}.b)--d)].
This experimental observation matches exactly the radius calculated based on 
\eq{t-ast}. Indeed, for the data shown in \Fig{DeltaT_n0-scaling} 
we find values for $r^\ast$ that decrease from $20\,\mu$m for small values of 
$\dichte \kappa^{-3} \xi^{-4}$ to $10\,\mu$m for the largest considered values.

It is instructive to compare $r^{\ast}$ to the droplet radius, $r_{\!\scriptsize\Pen}$, where the P\`eclet number, \Pen,
of the droplet motion crosses one.  Calculating 
$\Pen = u L/D_d$ based on 
the sedimentation velocity $u=2 \Delta\rho \, g \, r_{\!\scriptsize\Pen}^2 / (9\eta)$,
the droplet diameter $L=2 r_{\!\scriptsize\Pen}$, and the Brownian droplet diffusivity 
$D_d = k_B \, T / (6 \pi \eta \, r_{\!\scriptsize\Pen})$ 
yields
\begin{equation}
  \Pen = \frac{8\pi}{3} \; \frac{ \Delta\rho \: g }{ k_B \, T } \;  r_{\!\scriptsize\Pen}^4  \, .
\end{equation}
It takes values $\Pen \simeq 1$ for droplet radii
$r_{\!\scriptsize\Pen} 
=      \left[ k_B \, T / (8\pi \, \Delta\rho \: g) \right]^{1/4} 
\simeq 1.2 \,\mu$m.  
In line with expectation, the bottleneck radius $r^{\ast}$ is of the
same order of magnitude, even though somewhat larger than
$r_{\!\scriptsize\Pen}$.

\subsection{The role of Ostwald ripening}

\citet{Wilkinson2014} obtained \eq{Delta_t} based on an analysis of
the crossover from classical Ostwald ripening to the collection
scenario also adopted in the present theory. His derivation does not
provide a physical interpretation of the values of $\alpha$, and was
criticised in \citet{RohloffLappVollmer2014} for not predicting
physically sound values of $\Delta t$ and the bottleneck
radius,~$r^{\ast}$.

Classical Ostwald ripening is encountered for $k = 1$. For this value
\eq{def-alpha} does not apply because \eq{accretion-rate} is obtained
from the general equation~\eq{r-dot-frontier} as a large $k$ limit.
\EQ{r-dot-frontier} approaches the asymptotic scaling solution of
Ostwald ripening for $k=1$ where $0 \leq a/\Rav \leq 3/2$
\citep{LifshitzSlyozov1961}.  Consequently, the largest droplets
follow \eq{r-dot-frontier} with $k \simeq 3/2$, and \eq{Delta_t} is
recovered with a value $\alpha$ provided by \eq{def-alpha} evaluated
for $k=3/2$.  Hence, we find a value of $\alpha = 4.36$, which results
in a prediction of $\Delta t$ that is too large by a factor of about
six.

The error in the prediction of the bottleneck radius is even more severe. 
When evaluating the expression 
\begin{equation}
   r^\ast =\left(\frac{12 (k-1)^2 D^2 \sigma^2}{\collEfficiency \kappa \xi}\right)^{1/7}
\label{eq:bottleneck}
\end{equation} 
provided in \citet{Wilkinson2014} and \citet{RohloffLappVollmer2014},
one obtains values in the order of $0.1\,\mu$m that are too small by
two orders of magnitude. In particular, they are much smaller than the
value where $\Pen = 1$.  These discrepancies rule out Ostwald ripening
as a relevant contribution to growth in our experiments.
In this respect our findings are fully analogous with the description
of warm terrestrial rain where the effects of Ostwald ripening is also
believed to be insignificant \citep{Clement2008}.

\subsection{Predicting $\Delta t$ for warm terrestrial rain}
\label{sec:rain}

It is instructive to evaluate, \eq{DeltaT37}, for common
situations in warm rain \citep{BeardOchs1993,MoranMorgan1997}.  

The number density of droplets has been determined in recent
measurement campaigns \citep{DitasShawSiebertSimmelWehnerWiedensohler2012}, 
yielding $n \simeq 4.7\times 10^{8}$m$^{-3}$.  The material
constants entering the settling velocity of the droplets are the
density contrast of water and air,
$\Delta \rho \approx 10^3$kg/m$^3$, 
and the dynamic viscosity of air, 
$\mu = 1.8\times 10^{-5}$kg\,m$^{-1}$s$^{-1}$ at 10$^{\circ}$C
\cite[p.~103]{Rogers1989}. 
Given that the dynamic viscosity of air is much smaller than that of water, 
this provides a value
$\kappa = 1.2 \times 10^8$m$^{-1}$s$^{-1}$ [\cf\eq{u}].
Moreover, the ramp rates $\xi$ were estimated in
\citet{VollmerPapkeRohloff2014} to lie in the range 
$\xi = 5\times 10^{-6} \dots 5\times 10^{-5}$s$^{-1}$.
For a collection efficiency of $\collEfficiency = 0.3$, equation \Eq{DeltaT37}
then provides time scales $\Delta t$ in the range of $10\,$s and $30\,$s,
and bottleneck radii of the order of $30\,\mu$m.

The value observed for the bottleneck radius matches expectation
\citep{KostinskiShaw2005,Clement2008}.  On the other hand, the value of
$\Delta t$ is too small as compared to experiments.  This is extremely
remarkable, because common estimates
\citep{Houghton1959,falkovich02,Clement2008} based on diffusive ripening
processes and growth by collection tend to
provide estimates that are rather too large. Indeed, this is also what
one finds \citep{Wilkinson2014} when using \eq{Delta_t} with 
$\alpha \simeq 1$.
We attribute this discrepancy to limitations of the expression,
\eq{sweep-rate}, for the growth by collection.  A model that
only considers the size of the largest droplets tends to
overestimate the growth speed of the droplets in this regime. After all,
the collision frequency entering \Eq{sweep-rate} should be based on the 
relative droplet velocity
rather than on the falling velocity, \eq{u}, of the large
droplets. For the binary mixtures considered in the present paper 
the settling velocity of the small
droplets is negligible such that the approximation holds. However, for
systems with a large mass density contrast, $\Delta\rho$, \ie in particular
rain droplets in clouds, this is probably not justified. Follow-up
work is in progress, where we incorporate information on the evolution
of the full droplet distribution, in order to enhance the model to
also cover this case.

\section{Conclusion}
\label{sec:conclusion}

In the present paper we have established a faithful description of the
period, $\Delta t$, of episodic precipitation in binary mixtures.
It is based on a low-dimensional model accounting only for the
interplay of diffusive droplet growth and a runaway instability of the
droplet size that arises when the largest droplets start to be
effected by buoyancy.
The model neither accounts for spatial degrees of freedom, nor for the
droplet size distribution.
In contrast to systems featuring reactive flow, the disregarding of 
spatial degrees of freedom is justified: for the nonlinear
reactions terms that characterise phase separation, the convective mixing
efficiently eliminates spatial inhomogeneities of the droplet size
distribution \citep{BenczikVollmer2010,BenczikVollmer2012}.
In addition, detailed knowledge about the droplet size distribution is
not needed to predict $\Delta t$ as long as there is some
polydispersity in the distribution such that the largest droplets can
effectively grow by collecting small droplets.
The treatment of growth by collection has been inspired by models for
initiation of warm rain \citep{Houghton1959,BeardOchs1993,KostinskiShaw2005}.
However, in contrast to earlier work we modelled the diffusive growth
according to recently established models for aggregate growth in
the presence of a sustained ramping of the droplet volume fraction
\citep{ClarkKumarOwenChan2011,VollmerPapkeRohloff2014}.
Combining the impact of the resulting diffusive growth, that is most
effective for very small droplets, and growth by collection, that
arises when the largest droplets reach a size where their P\`eclet
number surpasses one, provides a low-dimensional model for the
dependence of $\Delta t$ on the number density of droplets, \dichte,
the ramp rate, $\xi$, the collection efficiency, and material
constants fixing the Stokes settling velocity of the droplets.
This results in a master plot, \Fig{DeltaT_n0-scaling}, where data
for various ramp rates and temperatures collapse on the
theoretical prediction \Eq{DeltaT37}. The only free parameter in this
fit is the collection efficiency that is expected to take values in the range
between $0.1$ and $1$ \citep{BeardOchs1993}.
The theory also provides a relation, \Eq{r-dot-frontier}, between the
droplet number density, \dichte, the ramp rate, $\xi$, and material
constants characterising diffusive droplet growth. This relation
can be used to eliminate \dichte\ from \Eq{DeltaT37}, thus obtaining a
prediction \eq{Delta_t} connecting $\xi^{3/7} \, \Delta t$ to a
nontrivial combination of material constants that is a known function
of temperature. 
The master plots shown in \Fig{master} demonstrate that this prediction is in
quantitative agreement with a vast set of data obtained for repeated
waves of precipitation in both phases of water/isobutoxyethanol and
methanol/hexane mixtures.
The data collapse establishes that the bottleneck of droplet growth
quantitatively determines the time scale, $\Delta t$, of rain
initiation in binary mixtures, and its parameter dependence.  The
bottleneck corresponds to the minimum of the droplet growth speed,
arising for intermediate droplet radii where growth by diffusive
collection of supersaturation is no longer effective, and collection
of smaller droplets by large sedimenting droplets is not yet effective
because buoyancy is still negligible. The time scale $\Delta t$
amounts to a small multiple of the time needed to cross this
bottleneck.

Follow-up work will address the evolution of the full droplet size
distribution in order to explore how to reconcile the tendency of the
distribution to become more monodisperse
\citep{KleinMoisar1963,Sugimoto1992,WallaceHobbsBook2006,ClarkKumarOwenChan2011,VollmerPapkeRohloff2014}
with the observation that growth of large droplets by
collecting smaller ones works best for a large size mismatch
\citep{WallaceHobbsBook2006}.  The excellent data collapse documented
in \Fig{DeltaT_n0-scaling} and \Fig{master} suggests that the
approximation to still consider the small droplets in the runaway
regime as Brownian particles seems to be well-justified for binary
mixtures.  In contrast, our estimate for warm terrestrial rain,
\Sect{rain}, suggests that our model predicts too small values for
$\Delta t$ due to an approximation of the droplet collection rates
that need not hold for terrestrial rain.
Extending the present work towards mixtures with a larger mass density
contrast will allow us to systematically develop models addressing the
emergence of precipitation in systems with a higher mass-density
contrast between the coexisting phases. In particular, these
generalisations of the model will allow us to address the growth of
droplets in terrestrial \citep{KostinskiShaw2005,GrabowskiWang2013}
and exo-planetary clouds \citep{MarleyAckermanCuzziKitzmann2013}.

\begin{acknowledgments}

  Our views on the theoretical interpretation of the present data
  developed in intensive discussions with Michael Wilkinson, who also
  proposed to denote the investigated repeated nucleation and
  sedimentation cycles as episodic precipitation.
  In addition, we acknowledge very useful discussion with Charles
  Clement, Izabella Benczik, Itzhak Fouxon, Raymond Pierrehumbert,
  Raymond Shaw, Axel Seifert, and Valerio Lucarini, and we are
  grateful to 
  Greg Bewley, 
  Stephan Herminghaus,
  Jakob de Maeyer, and 
  Marco Mazza
  for comments on the manuscript.

\end{acknowledgments}

\appendix

\section{Material Constants}
\label{sec:MatConst}

\begin{figure}
  \includegraphics[width=\textwidth]{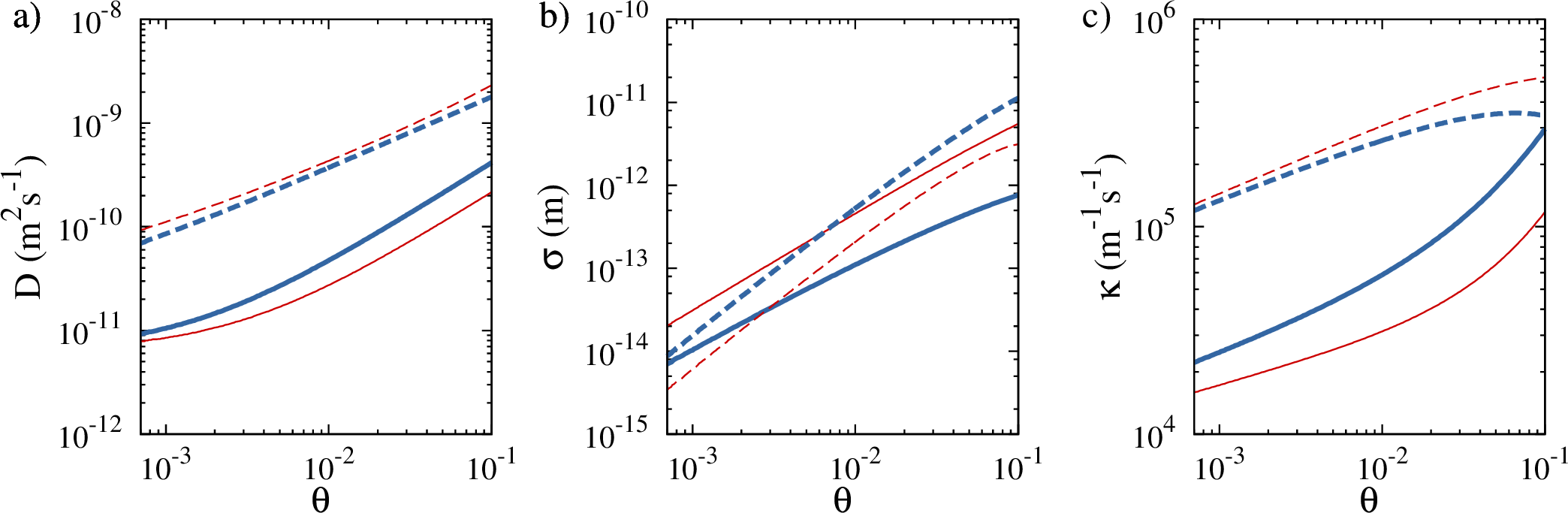}
  \caption[]{\textbf{Material constants}. 
    a) The diffusion coefficient $D$, 
    b) the Kelvin length $\sigma$, and 
    c) the sedimentation prefactor $\kappa$ as a function of the reduced
    temperature $\theta$ for IBE+W (solid lines) and M+H (dashed lines). 
    The thick blue and the thin red lines show the dependence in the lower 
    and the upper layer of the fluid mixtures, respectively.
  \label{fig:material_constants}}
\end{figure}

\FIG{orig-data} shows the period, $\Delta t$, of episodic precipitation for 
different ramp rates, $\xi$. Different data points for a given
ramp rate are due to the drift of $\Delta t$ when the pertinent
material constants, $D$, $\sigma$, and $\kappa(\Delta\rho, \visc_b, \visc_d )$
change upon moving further away from the critical point.
In the following we provide the temperature dependence of these material constants.
We cite the data here as they were provided in the original
literature (even when we are in doubt that they are
accurate to six significant digits for our samples). 
Upon doing so we denote the mass fraction as $\phi^m$ and the molar
fraction as $\phi^n$, respectively.
The resulting temperature dependence of the diffusion coefficient $D$,
the Kelvin length $\sigma$ provided by \eq{sigma}, and the
sedimentation prefactor $\kappa$ provided by \eq{u}, are summarised in
\Fig{material_constants} in order to give easy access to the constants
appearing in the predictions \Eq{DeltaT37} and \Eq{Delta_t}.
The temperature dependence translates to a time dependence when
inverting the protocol $T(t)$ of the temperature ramp.

\subsection{Isobutoxyethanol and water} 

The theoretical curves in \Figs{DeltaT_n0-scaling} and \fig{master} use data on material parameters from
a variety of sources \citep{Steinhoff1995,Ara+90,Doi2000,Menzel2003,Douheret2002}.
The index $i\in\{\textrm{IBE},\textrm{W}\}$ will be used to refer to
material properties of IBE and water, respectively, and in accordance
with the phase diagram, \Fig{phasdias}.a), the concentration
are always given in terms of $\phi = \phi_{\textrm{IBE}}$.

\subsubsection{Density  {\protect{\citep[based on][]{Doi2000}}}}

The densities of the phases are determined by the composition, thermal
expansion and molar excess volume,
\begin{equation}
  \rho(\phi^{m}, T) 
  = 
  \left[ \frac{\phi^{m}}{\rho_{\textrm{IBE}}} 
    + \frac{1-\phi^{m}}{\rho_{\textrm{W}}} 
    + \left( \frac{\phi^{m}}{M_{\textrm{IBE}}} 
      + \frac{1-\phi^{m}}{M_{\textrm{W}}}
    \right) V_{\rm E}^{n} \right]^{-1} \, ,
  \label{eq:rho4}
\end{equation}
where $\rho_{i} = \rho_i(T)$ are the (temperature-dependent) densities of the pure substances, $M_i$ their
molar masses, and $V_{\rm E}^{n} = V_{\rm E}^{n}(\phi^{n})$ is the molar excess volume. 

The molar masses, $M_i$ are $18.01528\,$g/mol for
water \citep{properties_water} and $118.17416\,$g/mol for IBE \citep{properties_IBE}, respectively.

The temperature dependence of the density, $\rho_{i}(T)$, 
of the pure substances is linearly approximated around $T_0 = 25^{\circ}$C,
\begin{equation}
 \rho_{i}(T) = \rho_{i}(T_{\rm 0}) - \alpha_{i} \; (T-T_{\rm 0})
  \label{eq:rho2}
\end{equation}
with fit parameters for $\rho_{i}$ and $\alpha_{i}$ given in \Tab{thermalexpansion}.
\begin{table}
	\centering
		\begin{tabular}{l@{$\qquad$}c@{\qquad}c}	
					               &	$\rho_i (T_{\rm 0})$ \  [g cm$^{-3}$]	& 	$\alpha_i$	\  [g cm$^{-3}$ K$^{-1}$] \\[2mm]
			water $\qquad$     & $0.997043$	                          &   $0.2571 \times 10^{-3}$              \\
			IBE   $\qquad$     & $0.886255$                           &   $0.968 \times 10^{-3}$	             
		\end{tabular}
		\caption{Densities and thermal expansion coefficients for water
      and IBE according to \citet{Doi2000}.}
		\label{tab:thermalexpansion}
\end{table}

Moreover, the molar excess volume is fitted according to \citet{Doi2000}:
\begin{eqnarray}
  V_{\rm E}^{n}(\phi^{n}) 
  &=& 
  \frac{\phi^{n} \; (1-\phi^{n})}{1-G\;\tilde\phi} 
  \;
  \left(
    A_{1} + A_{2} \;\tilde\phi + A_{3} \;\tilde\phi^2 
  \right)
  \label{eq:excessvolume1} 
  \\[2mm]
  \textrm{with } \quad
  \tilde\phi &=& 1-2\phi^{n}
  \nonumber
  \\[2mm]
  \textrm{and }  \quad
  G  &=&  0.975 \, , 
  \nonumber
  \\[2mm]
  A_{1}  &=&  -3.079 \; \textrm{cm$^{3}$/mol} \, , 
  \nonumber
  \\[2mm]
  A_{2}  &=&  1.801 \; \nobreak\mbox{cm$^{3}$/mol} \, , 
  \nonumber
  \\[2mm]
  A_{3}  &=&  0.839 \; \nobreak\mbox{cm$^{3}$/mol} \, .
  \nonumber
\end{eqnarray}
A slight temperature dependence of these fit parameters was reported
in \citet{Doi2000}. However, it is so small that we need not take it
into account here.

To get the dependence of the density difference on the reduced
temperature the dependence $\phi(\theta)$ (coexistence curve) 
into \eq{rho4}.

\subsubsection{Viscosity   \citep[own measurements augmented by data of][]{Weast1988,Menzel2003}}

We first provide the data of the pure phases, and then obtain the
viscosity of the mixture by appropriate interpolation.

Following \citet{Weast1988} we describe the temperature dependence of
the pure substances by
\begin{equation}
    \visc_i(T) = A_i \; 10^{\frac{B_i\: (T_0 - T) - C_i \: (T_0-T)^2}{T+D_i}} \, .
    \label{eq:viscosity1}
\end{equation}
In \Tab{coeffvisc} we provide the values for pure water
provided in \citet{Weast1988}, and parameters of a fit for IBE whose
viscosity we determined with an Ubbelohde viscometer type 537 10/I
made by Schott.

\begin{table}
      \centering
      \begin{tabular}{l@{$\qquad$}l@{$\quad$}l@{$\qquad$}l@{$\quad$}l}
				  & $A$ \ [kg\,m$^{-1}$ s$^{-1}$]	& $B$	& $C$ \ [($^\circ$ C)$^{-1}$]	& $D$ \	[$^\circ$ C]	\\[2mm]
	  water & $1.002\times 10^{-3}$ & 1.3272 & 0.001053 & 105\\ 
	  IBE  & $3.4\times 10^{-3}$  & 1.7  & 0.001    & 110
      \end{tabular}
      \caption{Data of \citet{Weast1988} for the fit coefficients for
        the viscosity of water and IBE, defined by 
        \eq{viscosity1}. In both cases $T_0 = 20^\circ$C is used as reference temperature.}
	  \label{tab:coeffvisc}
\end{table}

To interpolate the viscosities for a mixed phase of given mass
fraction $\phi^m$ we use the composition-dependent viscosities at
the reference temperature $T_r = 25^{\circ}$C for a homogeneous mixture in the single-phase
regime \citep{Menzel2003}. 
The data is fitted with a fifth order polynomial
\begin{subequations}
\begin{eqnarray}
		\visc(\phi^m, T_{r}) 
    &=& -40.66   \, (\phi^m)^5 
      + 103.44 \, (\phi^m)^4 
      - 100.32 \, (\phi^m)^3 
      \nonumber\\
      && \qquad	
      + 39.35  \, (\phi^m)^2 
      + 0.17   \, \phi^m 
      + 0.91
  \label{eq:viscosity2}
  \\
  &=& w_{\visc}(\phi^m) \; \visc_{\textrm{IBE}}(T = 25^{\circ}{\rm C}) 
  + \left[ 1 - w_{\visc}(\phi^m) \right] \: \visc_{\textrm{W}}(T = 25^{\circ}{\rm C})
  \, ,
  \label{eq:visc-interpolation}
\end{eqnarray}%
\end{subequations}%
where the latter equation \emph{defines} the dimensionless, non-linear weight
function $w_\visc$ that expresses $\visc(\phi^m, T = 25^{\circ}{\rm C})$
as a function of the viscosities of the pure substances
$\visc_{\textrm{IBE}}(T = 25^{\circ}{\rm C})$ and 
$\visc_{\textrm{W}}(T = 25^{\circ}{\rm C})$, respectively. 

Assuming that this weight function is not varying 
substantially in the temperature range of our measurements, we can
use the interpolation \eq{visc-interpolation} to determine the
viscosity of the mixture also at other temperatures. After all, the
temperature dependence $\visc_{\textrm{IBE}}(T)$ and
$\visc_{\textrm{W}}(T)$ were provided by \eq{viscosity1} with coefficients
in \Tab{coeffvisc}.
To check the strong assumption entering this interpolation, 
\begin{equation}
  \visc(\phi^m,T) 
  = w_{\visc}(\phi^m) \; \visc_{\textrm{IBE}}(T) 
  + \left[ 1 - w_{\visc}(\phi^m) \right] \; \visc_{\textrm{W}}(T)
  \, ,
\label{eq:viscosity3}
\end{equation}
we measured the viscosity of the two coexisting phases at
$T=40^{\circ}$C.  For both phases the prediction of \eq{viscosity3}
was accurate to within 2\%. This is sufficient for our purposes.

\subsubsection{Diffusion coefficient \citep[based on][]{Steinhoff1995}}

The renormalisation group theory predicts that the diffusion
coefficient vanishes when the critical point is approached.  On the
other hand, the renormalisation group theory is precise only in the
vicinity of the critical point, and its application to interdiffusion
coefficients has been a source of
controversy \citep{Sengers1985,DasFisherSengersHorbachBinder2006}. 
For this reason we choose to rely upon interpolations of experimental data.
We do not expect that our data follow the critical exponents because 
the temperatures in our experiments lie outside the critical region. 
Hence, we fitted the data of \citet{Steinhoff1995} with the following
expression:
\begin{equation}
  D_i(\theta) = D_{\rm c} + \delta_i \; \theta 
  \label{eq:diffusioncoefficient}
\end{equation}
with 
$D_{\rm c} = 6.4\times 10^{-12}$m$^{2}$/s,
$\delta_{\textrm{IBE}} = 2.1\times 10^{-9}$m$^{2}$/s and
$\delta_{\textrm{W}} = 4.1\times 10^{-9}$m$^{2}$/s.

\subsubsection{Interfacial tension \citep[based on][]{Ara+90}}

The interfacial tension vanishes at the critical temperature, and its
dependence at higher temperatures can be represented by a power law
\begin{equation}
 \gamma(\theta) = \gamma_{0} \: \theta^{\alpha_{\gamma}} \, ,
  \label{eq:interfacialtension}
\end{equation}
where a fit to the data of \citet{Ara+90} yields
$\gamma_{0} = 7.3\times 10^{-4}$N/m and $\alpha_{\gamma} = 1.2$.

\subsubsection{Molar volume \citep[based on][]{Douheret2002}}

According to \citet{Douheret2002} the molar volume $V^n$ can be approximated by
\begin{equation}
	V^n 
  = \phi^{n}     \; V^n_{\textrm{IBE}} 
  + (1-\phi^{n}) \; V^n_{\textrm{W}}
	\label{eq:molarvolume}
\end{equation}
with $V^n_{\textrm{IBE}} = 124$cm$^{3}$/mol and $V^n_{\textrm{W}} = 15.98$cm$^{3}$/mol.

\subsection{Methanol/hexane mixtures}

In this subsection the index $i\in\{M,H\}$ denotes material constants
of the methanol and hexane, respectively, and concentrations refer to
methanol, $\phi = \phi_{\textrm{M}}$.

\subsubsection{Density \citep[based on][]{AbbasSatherleyPenfold1997,Orge1997}}

The densities are again calculated according to \eq{rho4}.
In this case the molar mass is $32.04186\,$g/mol for methanol \citep{properties_methanol} and
$86.17536\,$g/mol for hexane \citep{properties_hexane}.
The temperature dependence of the pure substances amounts to \citep{AbbasSatherleyPenfold1997}
\begin{equation}
 \rho_i(T) = a_0 + a_1\: T +a_2 \: T^2
  \label{eq:rho3}
\end{equation}
with coefficients given in \Tab{methexdensity}.
\begin{table}
  \centering
  \begin{tabular}{l@{$\qquad$}l@{$\quad$}l@{$\quad$}l} 
             & $a_0$ [g cm$^{-3}$]& $a_1$ [g cm$^{-3}$ K$^{-1}$] & $a_2$ [g cm$^{-3}$ K$^{-2}$]	\\[2mm] 
    methanol & 1.382             & $-3.135 \times 10^{-3}$    & $3.813\times 10^{-6}$ \\ 
    hexane   & 0.6839            & $6.989  \times 10^{-4}$    & $-2.656\times 10^{-6}$
  \end{tabular}
  \caption{Coefficients of the density \citep{AbbasSatherleyPenfold1997}.}
  \label{tab:methexdensity}
\end{table}
The excess volume is expressed as \citep{Orge1997}
\begin{subequations}
\begin{eqnarray}
  V_E^n (\phi^n) &=& \phi^n \; (1-\phi^n) \; 
  \left[ B_0 + B_1 \tilde\phi + B_2 \tilde\phi^2 \right]
\\[2mm]
\mbox{with } \quad
  \tilde\phi &=& 1-2\phi^{n}
  \nonumber
  \\[2mm]
  \mbox{and }  \quad
B_0 &=& 2.0741 \; \nobreak\mbox{cm$^3$/mol}, \nonumber \\
B_1 &=& 0.3195 \; \nobreak\mbox{cm$^3$/mol}, \nonumber \\
B_2 &=& 1.7733 \; \nobreak\mbox{cm$^3$/mol}. \nonumber
\end{eqnarray}
\end{subequations}

\subsubsection{Viscosity \citep[based on][]{Assael1994,Eicher1972,Orge1997}}

We first provide the data of the pure phases, and then obtain the
viscosity of the mixture by appropriate interpolation.

The viscosity of pure methanol \citep{Assael1994} is
\begin{equation}
 \visc_{\textrm{M}}(T) = A \: \exp(B/T)
\end{equation}
with $A = 8.203 \times 10^{-6}$Pa s and $B = 1251.4\,$K. 

For hexane our analysis is based on the kinematic viscosity 
$\nu_{\textrm{H}}$ provided in \citet{Eicher1972} 
\begin{equation}
 \nu_{\textrm{H}}(T) = \nu' \left(\frac{T}{T'}\right)^n \exp \left(  \frac{B(T'-T)}{(T'-T_0)(T-T_0)}\right)
\end{equation} 
with $n=-2.24057$, $B=4.78496\,$K and $T_0 = 222.468\,$K, 
reference viscosity $\nu' = 0.4604 \times 10^{-10}$m$^2$/s, and 
reference temperature $T'= 296.267\,$K. 
Together with the density of hexane, which is provided in \eq{rho3},
this provides the dynamic viscosity 
$\visc_{\textrm{H}} = \rho_{\textrm{H}} \: \nu_{\textrm{H}}$.

The viscosity of the mixture is obtained by interpolating based on the excess
viscosity provided in \citet{Orge1997}
\begin{eqnarray}
 \visc(  \phi^n,T  )
& = & \phi^n \, \visc_{\textrm{M}}(T) 
 + (1-\phi^n) \, \visc_{\textrm{H}}(T) 
+ \, \phi^n \; (1-\phi^n) \, \left[ B_0 + B_1 (1-2\,\phi^n) \right] \quad
\\[2mm]
\mbox{with} \quad B_0 &=& -1.83\times 10^{-4} \nobreak\mbox{kg\,m$^{-1}$\,s$^{-1}$}
\nonumber \\[2mm]
B_1 & = & 0.91\times 10^{-4} \nobreak\mbox{kg\,m$^{-1}$\,s$^{-1}$} \, .
\nonumber
\end{eqnarray}

\subsubsection{Diffusion coefficient  \citep[based on][]{Clark1986}}

The dependence of the diffusion coefficient $D(\phi^n, \theta)$ on the
concentration $\phi^n$ of the mixture and on the reduced temperature
$\theta$ can be approximated by \citep{Clark1986}
\begin{eqnarray}
  D(\phi^n, \theta) 
  &=&  A_0 
  + A_1 \: \phi^n 
  + A_2 \: (\phi^n)^2
  + A_3 \: (\phi^n)^3 
  + A_4 \: (\phi^n)^4 
  + A_\theta \: \theta^{0.68516}
\end{eqnarray}
\begin{eqnarray}
\mbox{with } \qquad
A_0  &=&  3.2457\times 10^{-9}\,\nobreak\mbox{m$^2$/s}, 
\nonumber\\[2mm]
A_1  &=&  -1.68497\times 10^{-8}\,\nobreak\mbox{m$^2$/s}, 
\nonumber\\[2mm]
A_2  &=&  3.63103\times 10^{-8}\,\nobreak\mbox{m$^2$/s}, 
\nonumber\\[2mm]
A_3  &=&  -4.1949\times 10^{-8}\,\nobreak\mbox{m$^2$/s}, 
\nonumber\\[2mm]
A_4  &=&  2.223\times 10^{-8}\,\nobreak\mbox{m$^2$/s}, 
\nonumber\\[4mm]
\nobreak\mbox{and } \qquad
A_\theta  &=&  2.5067\times 10^{-9}\, \nobreak\mbox{m$^2$/s} \, .
\end{eqnarray}
Similarly to the expression
\eq{diffusioncoefficient} the fit for the M+H mixture 
involves a constant background contribution, and the singular contribution expected from the theory of critical phenomena. 
\citet{Clark1986} fitted the composition dependence of the
background contribution by a forth-order
polynomial in $\phi^n$, and introduced the term $A_5 \:
\theta^{0.68516}$ to account for the singular contribution to the
diffusion. The latter term vanishes at $T=T_c$ with the appropriate critical scaling
exponent, $0.68516$.  

\subsubsection{Interfacial tension \citep[according to][]{AbbasSatherleyPenfold1997}}

Data of interfacial tension \citep{AbbasSatherleyPenfold1997} are parametrised according to \eq{interfacialtension} with $\gamma_{0} =
3.631\times 10^{-2}$N/m and $\alpha_{\gamma} = 1.65$. This data lies
beyond the critical region of $\theta < 10^{-2.5}$ where scaling with
a critical exponents is expected \citep{AbbasSatherleyPenfold1997}.

\subsubsection{Molar volume \citep[according to][]{Maruyama1995}}

The molar volume is interpolated with \eq{molarvolume}
with $V^n = 41.1$cm$^{3}$/mol for methanol and $V^n =
133.2$cm$^{3}$/mol for hexane \citep{Maruyama1995}.


\begin{thebibliography}{86}
\expandafter\ifx\csname natexlab\endcsname\relax\def\natexlab#1{#1}\fi

\bibitem[Aarts {\em et~al.\/}(2005)Aarts, Dullens \&
  Lekkerkerker]{AartsDullensLekkerkerker2005}
{\sc Aarts, D.G., Dullens, R.P.A. \& Lekkerkerker, H.N.W.} 2005 Interfacial
  dynamics in demixing systems with ultralow interfacial tension. {\em New J.
  Phys.\/} {\bf 7}, 40.

\bibitem[Abbas {\em et~al.\/}(1997)Abbas, Satherley \&
  Penfold]{AbbasSatherleyPenfold1997}
{\sc Abbas, Shabira, Satherley, John \& Penfold, Robert} 1997 The liquid-liquid
  coexistence curve and the interfacial tension of the methanol-n-hexane
  system. {\em J. Chem. Soc. Farad Trans.\/} {\bf 93}, 2083--2089.

\bibitem[Aizpiri {\em et~al.\/}(1990)Aizpiri, Correa, Rubio \&
  {Pe\~{n}a}]{AizpiriCorreaRubioPena1990}
{\sc Aizpiri, Arturo~G., Correa, {Jos\'{e} A.}, Rubio, {Ram\'{o}n G.} \&
  {Pe\~{n}a}, {Mateo Dr\v{i}az}} 1990 Coexistence curve of methanol+n-heptane:
  {Range} of simple scaling and critical amplitudes. {\em Phys. Rev. B\/} {\bf
  41}, 9003.

\bibitem[Aratono {\em et~al.\/}(1990)Aratono, Nakayama, Ikeda \&
  Motomura]{Ara+90}
{\sc Aratono, M., Nakayama, S., Ikeda, N. \& Motomura, K.} 1990 Thermodynamic
  consideration on the interface formation of water and ethylene glycol
  isobutyl ether mixture. {\em Coll. Polymer Sc.\/} {\bf 268}, 877--82.

\bibitem[Assael \& Polimatidou(1994)]{Assael1994}
{\sc Assael, M.~J. \& Polimatidou, S.~K.} 1994 Measurements of the viscosity of
  alcohols in the temperature range {$290$--$340\,$K} at pressures up to
  {$30\,$MPa}. {\em Int. J. Thermophys.\/} {\bf 15}~(1), 95--107.

\bibitem[Auernhammer {\em et~al.\/}(2005)Auernhammer, Vollmer \&
  Vollmer]{auernhammer05JCP}
{\sc Auernhammer, G\"unter~K., Vollmer, Doris \& Vollmer, J\"urgen} 2005
  Oscillatory instabilities in phase separation of binary mixtures: Fixing the
  thermodynamic driving. {\em J. Chem. Phys.\/} {\bf 123}, 134511.

\bibitem[Beard \& Ochs(1993)]{BeardOchs1993}
{\sc Beard, Kenneth~V. \& Ochs, Harry~T.} 1993 Warm-rain initiation: An
  overview of microphysical mechanisms. {\em J. Appl. Meteor.\/} {\bf 32},
  608--625.

\bibitem[Benczik \& Vollmer(2010)]{BenczikVollmer2010}
{\sc Benczik, I.J. \& Vollmer, J.} 2010 A reactive-flow model of phase
  separation in fluid binary mixtures with continuously ramped temperature.
  {\em EPL\/} {\bf 91}, 36003.

\bibitem[Benczik \& Vollmer(2012)]{BenczikVollmer2012}
{\sc Benczik, I.J. \& Vollmer, J.} 2012 A diffusion-induced transition in the
  phase separation of binary fluidmixtures subjected to a temperature ramp.
  {\em EPL\/} {\bf 100}~(1), 16001.

\bibitem[Beysens {\em et~al.\/}(1988)Beysens, Guenoun \&
  Perrot]{BeysensGuenounPerrot1988}
{\sc Beysens, D., Guenoun, P. \& Perrot, F.} 1988 Phase separation of critical
  binary fluids under microgravity: Comparison with matched-density conditions.
  {\em Phys. Rev. A\/} {\bf 38}, 4173--4185.

\bibitem[Binder \& Stauffer(1976)]{BinderStauffer1976}
{\sc Binder, K \& Stauffer, D} 1976 Statistical theory of nucleation,
  condensation and coagulation. {\em Adv. Phys.\/} {\bf 25}, 343--396.

\bibitem[Blyth {\em et~al.\/}(2013)Blyth, Lowenstein, Huang, Cui, Davies \&
  Carslaw]{BlythLowensteinHuangCuiDaviesCarslaw2013}
{\sc Blyth, Alan~M., Lowenstein, Jason~H., Huang, Yahui, Cui, Zhiqiang, Davies,
  Stewart \& Carslaw, Kenneth~S.} 2013 The production of warm rain in shallow
  maritime cumulus clouds. {\em Quart. J. Roy. Met. Soc.\/} {\bf 139}, 20--31.

\bibitem[Bray(1994)]{Bray1994}
{\sc Bray, Alan~J.} 1994 Theory of phase-ordering kinetics. {\em Adv. Phys.\/}
  {\bf 43}, 357 -- 459.

\bibitem[Cashman \& Sparks(2013)]{CashmanSparks2013}
{\sc Cashman, Katharine~V. \& Sparks, R. Stephen~J.} 2013 How volcanoes work: A
  25 year perspective. {\em Geol. Soc. Am. Bull.\/} {\bf 125}~(5--6), 664--690.

\bibitem[Cates {\em et~al.\/}(2003)Cates, Vollmer, Wagner \&
  Vollmer]{cates03PhilTrans}
{\sc Cates, Michael~E., Vollmer, J\"urgen, Wagner, Alexander \& Vollmer, Doris}
  2003 Phase separation in binary fluid mixtures with continuously ramped
  temperature. {\em Phil. Trans. Roy. Soc. (Lond.) Ser. A\/} {\bf 361},
  793--807.

\bibitem[Cau \& Lacelle(1993)]{CauLacelle1993}
{\sc Cau, Franco \& Lacelle, Serge} 1993 Late-stage phase separation and
  sedimentation in a binary liquid mixture. {\em Phys. Rev. E\/} {\bf 47},
  1429--1432.

\bibitem[Clark {\em et~al.\/}(2011)Clark, Kumar, Owen \&
  Chan]{ClarkKumarOwenChan2011}
{\sc Clark, Michael~D., Kumar, Sanat~K., Owen, Jonathan~S. \& Chan, Emory~M.}
  2011 Focusing nanocrystal size distributions via production control. {\em
  Nano Lett.\/} {\bf 11}, 1976--1980.

\bibitem[Clark \& Rowley(1986)]{Clark1986}
{\sc Clark, W.~M. \& Rowley, R.~L.} 1986 The mutual diffusion coefficient of
  methanol-n-hexane near the consolute point. {\em AIChE Journal\/} {\bf
  32}~(7), 1125--1131.

\bibitem[Clement(2008)]{Clement2008}
{\sc Clement, Charles~F.} 2008 {\em Environmental Chemistry of Aerosols\/},
  chap. Mass Transfer to Aerosols, pp. 49--89. Oxford: Blackwell Publishing.

\bibitem[Das {\em et~al.\/}(2006)Das, Fisher, Sengers, Horbach \&
  Binder]{DasFisherSengersHorbachBinder2006}
{\sc Das, Subir~K., Fisher, Michael~E., Sengers, Jan~V., Horbach, J\"urgen \&
  Binder, Kurt} 2006 Critical dynamics in a binary fluid: Simulations and
  finite-size scaling. {\em Phys. Rev. Lett.\/} {\bf 97}, 025702.

\bibitem[Ditas {\em et~al.\/}(2012)Ditas, Shaw, Siebert, Simmel, Wehner \&
  Wiedensohler]{DitasShawSiebertSimmelWehnerWiedensohler2012}
{\sc Ditas, Florian, Shaw, Raymond~A., Siebert, Holger, Simmel, Martin, Wehner,
  Birgit \& Wiedensohler, Alfred} 2012 Aerosols-cloud
  microphysics-thermodynamics-turbulence: evaluating supersaturation in a
  marine stratocumulus cloud. {\em Atm. Chem Phys.\/} {\bf 12}~(5), 2459--2468.

\bibitem[Doi {\em et~al.\/}(2000)Doi, Tamura \& Murakami]{Doi2000}
{\sc Doi, Hideshige, Tamura, Katsutoshi \& Murakami, Sachio} 2000
  {Thermodynamic properties of aqueous solution of 2-isobutoxyethanol at T = 
  (293.15, 298.15, and 303.15) K, below and above LCST}. {\em J. Chem.
  Thermodyn.\/} {\bf 32}~(6), 729--741.

\bibitem[Douheret {\em et~al.\/}(2002)Douheret, Davis, Reis, Fjellanger, Vaage
  \& Hoiland]{Douheret2002}
{\sc Douheret, Gerard, Davis, Michael~I., Reis, Joao Carlos~R., Fjellanger,
  Inger~Johanne, Vaage, Marit~Bo \& Hoiland, Harald} 2002 Aggregative processes
  in aqueous solutions of isomeric 2-butoxyethanols at {$298.15\,$K}. {\em
  Phys. Chem. Chem. Phys.\/} {\bf 4}~(24), 6034--6042.

\bibitem[Eicher \& Zwolinski(1972)]{Eicher1972}
{\sc Eicher, Lawrence~D. \& Zwolinski, Bruno~J.} 1972 Molecular structure and
  shear viscosity. isomeric hexanes. {\em J. Phys. Chem.\/} {\bf 76}~(22),
  3295--3300.

\bibitem[Emmanuel \& Berkowitz(2006)]{EmmanuelBerkowitz2006}
{\sc Emmanuel, Simon \& Berkowitz, Brian} 2006 An experimental analogue for
  convection and phase separation in hydrothermal systems. {\em J. Geophys.
  Res.\/} {\bf 111}, B09103.

\bibitem[Falkovich {\em et~al.\/}(2002)Falkovich, Fouxon \&
  Stepanov]{falkovich02}
{\sc Falkovich, G., Fouxon, A. \& Stepanov, M.G.} 2002 Acceleration of rain
  initiation by cloud turbulence. {\em Nature\/} {\bf 419}, 151--154.

\bibitem[Farjoun \& Neu(2011)]{FarjounNeu2011}
{\sc Farjoun, Yossi \& Neu, John~C.} 2011 Aggregation according to classical
  kinetics: From nucleation to coarsening. {\em Phys. Rev. E\/} {\bf 83},
  051607.

\bibitem[Grabowski \& Wang(2013)]{GrabowskiWang2013}
{\sc Grabowski, Wojciech~W. \& Wang, Lian-Ping} 2013 Growth of cloud droplets
  in a turbulent environment. {\em Ann. Rev. Fluid Mech.\/} {\bf 45}, 293--324.

\bibitem[Guyon {\em et~al.\/}(2001)Guyon, Hulin, Petit \& Mitescu]{guyonBook}
{\sc Guyon, Etienne, Hulin, Jean-Pierre, Petit, Luc \& Mitescu, Catalin~D.}
  2001 {\em Physical Hydrodynamics\/}. Oxford: Oxford Univ. Press, translation
  from French: `Hydrodynamique Physique', 1991.

\bibitem[Han {\em et~al.\/}(2013)Han, Lu, McPherson, Keating, Moore, Park,
  Watson \& Jung]{HanLuMcPhersonKeatingMooreEtAl2013}
{\sc Han, Weon~S., Lu, M., McPherson, B.~J., Keating, E.~H., Moore, J., Park,
  E., Watson, Z.~T. \& Jung, N.-H.} 2013 Characteristics of {CO}$_2$-driven
  cold-water geyser, crystal geyser in {Utah}: experimental observation and
  mechanism analyses. {\em Geofluids\/} {\bf 13}~(3), 283--297.

\bibitem[Houghton(1959)]{Houghton1959}
{\sc Houghton, H.~G.} 1959 Cloud physics. {\em Science\/} {\bf 129}, 307--313.

\bibitem[Huang {\em et~al.\/}(1974)Huang, Goldburg \& Bjerkaas]{huang74}
{\sc Huang, John~S., Goldburg, Walter~I. \& Bjerkaas, Allan~W.} 1974 Study of
  phase separation in a critical binary liquid mixture: spinodal decomposition.
  {\em Phys. Rev. Lett.\/} {\bf 32}, 921--923.

\bibitem[Ingebritsen \& Rojstaczer(1993)]{IngebritsenRojstaczer1993}
{\sc Ingebritsen, S.~E. \& Rojstaczer, S.~A.} 1993 Controls on geyser
  periodicity. {\em Science\/} {\bf 262}~(5135), 889--92.

\bibitem[Iwanowski {\em et~al.\/}(2006)Iwanowski, Sattarow, Behrends, Mirzaev
  \& Kaatze]{IwanowskiSattarowBehrendsMirzaevKaatze2006}
{\sc Iwanowski, I., Sattarow, A., Behrends, R., Mirzaev, S.~Z. \& Kaatze, U.}
  2006 Dynamic scaling of the critical binary mixture methanol-hexane. {\em J.
  Chem. Phys.\/} {\bf 124}, 144505.

\bibitem[Kalwarczyk {\em et~al.\/}(2008)Kalwarczyk, Ziebacz, Fialkowski \&
  Holyst]{KalwarczykZiebaczFialkowskiHolyst2008}
{\sc Kalwarczyk, Tomasz, Ziebacz, Natalia, Fialkowski, Marcin \& Holyst,
  Robert} 2008 Late stage of the phase-separation process: Coalescence-induced
  coalescence, gravitational sedimentation, and collective evaporation
  mechanisms. {\em Langmuir\/} {\bf 24}, 6433 -- 6440.

\bibitem[Klein \& Moisar(1963)]{KleinMoisar1963}
{\sc Klein, E. \& Moisar, E.} 1963 {Elektronenmikroskopische und
  nephelometrische Untersuchungen \"uber das Kornwachstum von
  Silberhalogenidkristallen}. {\em Berichte Bunsenges. phys. Chem.\/} {\bf
  67}~(4), 349--355.

\bibitem[Kostinski \& Shaw(2005)]{KostinskiShaw2005}
{\sc Kostinski, Alexander~B. \& Shaw, Raymond~A.} 2005 Fluctuations and luck in
  droplet growth by coalescence. {\em Bull. Am. Met. Soc.\/} {\bf 86},
  235--244.

\bibitem[Koyaguchi {\em et~al.\/}(1990)Koyaguchi, Hallworth, Huppert \&
  Sparks]{KoyaguchiHallworthHuppertSparks1990}
{\sc Koyaguchi, Takehiro, Hallworth, Mark~A., Huppert, Herbert~E. \& Sparks, R.
  Stephen~J.} 1990 Sedimentation of particles from a convecting fluid. {\em
  Nature\/} {\bf 343}, 447 -- 450.

\bibitem[Lapp {\em et~al.\/}(2012)Lapp, Rohloff, Vollmer \&
  Hof]{LappRohloffVollmerHof2012}
{\sc Lapp, Tobias, Rohloff, Martin, Vollmer, J\"urgen \& Hof, Bj\"orn} 2012
  Particle tracking for polydisperse sedimenting droplets in phase separation.
  {\em Exp. Fluids\/} {\bf 52}, 1187--1200.

\bibitem[Leubner(2000)]{Leubner2000}
{\sc Leubner, Ingo~H.} 2000 Particle nucleation and growth models. {\em Curr.
  Opinion Coll. Interf. Sc.\/} {\bf 5}, 151--159.

\bibitem[Lifshitz \& Pitaevskii(1981)]{Lif+81}
{\sc Lifshitz, E.~M. \& Pitaevskii, L.~P.} 1981 {\em Landau and Lifshitz Course
  of Theoretical Physics, Vol.10 : Physical Kinetics\/}. Oxford:
  Butterworth-Heimemann.

\bibitem[Lifshitz \& Slyozov(1961)]{LifshitzSlyozov1961}
{\sc Lifshitz, Ilya~M. \& Slyozov, Vitaly~V.} 1961 The kinetics of
  precipitation from supersaturated solid solutions. {\em J. Phys. Chem.
  Solids\/} {\bf 19}~(1--2), 35 -- 50.

\bibitem[Marley {\em et~al.\/}(2013)Marley, Ackerman, Cuzzi \&
  Kitzmann]{MarleyAckermanCuzziKitzmann2013}
{\sc Marley, M., Ackerman, A., Cuzzi, J. \& Kitzmann, D.} 2013 Clouds and hazes
  in exoplanet atmospheres. In {\em Comparative Climatology of Terrestrial
  Planets\/} (ed. S.J. Mackwell, A.A. Simon-Miller, J.W. Harder \& M.A.
  Bullock), pp. 367--391. University of Arizona Press.

\bibitem[Martin \& Nokes(1988)]{MartinNokes1988}
{\sc Martin, Daniel \& Nokes, Roger} 1988 Crystal settling in a vigorously
  converting magma chamber. {\em Nature\/} {\bf 332}, 534 -- 536.

\bibitem[Maruyama {\em et~al.\/}(1995)Maruyama, Kawase, Tamaki \&
  Okazaki]{Maruyama1995}
{\sc Maruyama, K., Kawase, S., Tamaki, S. \& Okazaki, H.} 1995 Thermodynamic
  aspects of the {Rayleigh} and {Brillouin} scattering from a binary liquid
  mixture: {The} hexane-methanol system. {\em J. Phys. Chem.\/} {\bf 99}~(26),
  10644--10647.

\bibitem[McGraw \& Liu(2003)]{GrawLiu2003}
{\sc McGraw, Robert \& Liu, Yangang} 2003 Kinetic potential and barrier
  crossing: A model for warm cloud drizzle formation. {\em Phys. Rev. Lett.\/}
  {\bf 90}, 018501.

\bibitem[Menzel {\em et~al.\/}(2003)Menzel, Mirzaev \& Kaatze]{Menzel2003}
{\sc Menzel, K., Mirzaev, S.~Z. \& Kaatze, U.} 2003 Crossover behavior in
  micellar solutions with lower critical demixing point: Broadband ultrasonic
  spectrometry of the isobutoxyethanol-water system. {\em Phys. Rev. E\/} {\bf
  68}~(1), 011501.

\bibitem[Mirzaev {\em et~al.\/}(2010)Mirzaev, Heimburg \&
  Kaatze]{MirzaevHeimburgKaatze2010}
{\sc Mirzaev, Sirojiddin~Z., Heimburg, Thomas \& Kaatze, Udo} 2010 Critical
  behavior of polystyrene-cyclohexane: Heat capacity and mass density. {\em
  Phys. Rev. E\/} {\bf 82}, 061502.

\bibitem[Moran \& Morgan(1997)]{MoranMorgan1997}
{\sc Moran, Joseph~H. \& Morgan, Michael~D.} 1997 {\em Meteorolgy: The
  Atmosphere and the Science of Weather\/}, 5th edn. Upper Saddle River, NJ:
  Prentice-Hall.

\bibitem[Nakata {\em et~al.\/}(1982)Nakata, Dobashi, Kuwahara \&
  Kaneko]{NakataDobashiKuwaharaKaneko1982}
{\sc Nakata, Mitsuo, Dobashi, Toshiaki, Kuwahara, Nobuhiro \& Kaneko, Motozo}
  1982 Coexistence curve and diameter of the system ethylene glycol
  mono-isobutyl ether + water. {\em J. Chem. Soc. Farad Trans.\/} {\bf 78},
  1801--1810.

\bibitem[Nozawa {\em et~al.\/}(2005)Nozawa, Delville, Ushiki, Panizza \&
  Delville]{NozawaDelvilleUshikiPanizzaDelville2005}
{\sc Nozawa, Koh, Delville, Marie-H\'el\`ene, Ushiki, Hideharu, Panizza, Pascal
  \& Delville, Jean-Pierre} 2005 Growth of monodisperse mesoscopic metal-oxide
  colloids under constant monomer supply. {\em Phys. Rev. E\/} {\bf 72},
  011404.

\bibitem[Orge {\em et~al.\/}(1997)Orge, Iglesias, Rodríguez, Canosa \&
  Tojo]{Orge1997}
{\sc Orge, B., Iglesias, M., Rodríguez, A., Canosa, J.~M. \& Tojo, J.} 1997
  Mixing properties of (methanol, ethanol, or 1-propanol) with (n-pentane,
  n-hexane, n-heptane and n-octane) at {$298.15\,$K}. {\em Fluid Phase
  Equilibria\/} {\bf 133}~(1-2), 213--227.

\bibitem[PubChem(2013{\natexlab{{\em a\/}}})]{properties_water}
{\sc PubChem, {National Center for Biotechnology Information}}
  2013{\natexlab{{\em a\/}}} Pubchem compound database. CID=962 (17 July 2013).

\bibitem[PubChem(2013{\natexlab{{\em b\/}}})]{properties_IBE}
{\sc PubChem, {National Center for Biotechnology Information}}
  2013{\natexlab{{\em b\/}}} Pubchem compound database. CID=521158 (17 July
  2013).

\bibitem[PubChem(2013{\natexlab{{\em c\/}}})]{properties_methanol}
{\sc PubChem, {National Center for Biotechnology Information}}
  2013{\natexlab{{\em c\/}}} Pubchem compound database. CID=887 (17 July 2013).

\bibitem[PubChem(2013{\natexlab{{\em d\/}}})]{properties_hexane}
{\sc PubChem, {National Center for Biotechnology Information}}
  2013{\natexlab{{\em d\/}}} Pubchem compound database. CID=8058 (17 July
  2013).

\bibitem[Rimbert {\em et~al.\/}(2014)Rimbert, Claudotte, Gardin \&
  Lehmann]{RimbertClaudotteGardinLehmann2014}
{\sc Rimbert, N., Claudotte, L., Gardin, P. \& Lehmann, J.} 2014 Modeling the
  dynamics of precipitation and agglomeration of oxide inclusions in liquid
  steel. {\em Ind. Eng. Chem. Res.\/} {\bf 53}~(20), 8630--8639.

\bibitem[Rogers \& Yau(1989)]{Rogers1989}
{\sc Rogers, Roddy~R. \& Yau, M.~K.} 1989 {\em A Short Course in Cloud
  Physics\/}, 3rd edn., {\em International Series in natural philosophy\/},
  vol. 113. Pergamon Press, Oxford.

\bibitem[Rohloff {\em et~al.\/}(2014)Rohloff, Lapp \&
  Vollmer]{RohloffLappVollmer2014}
{\sc Rohloff, Martin, Lapp, Tobias \& Vollmer, J\"urgen} 2014 Comment on ``{A
  test-tube model for rainfall}'' by {Wilkinson Michael}. {\em EPL\/} {\bf
  108}, 30005.

\bibitem[Sam {\em et~al.\/}(2011)Sam, Hayase, Auernhammer \&
  Vollmer]{SamAuernhammerVollmer2011}
{\sc Sam, Ebie~M., Hayase, Yumino, Auernhammer, G\"unter~K. \& Vollmer, Doris}
  2011 Pattern formation in phase separating binary mixtures. {\em Phys. Chem.
  Chem. Phys.\/} {\bf 13}, 13333 -- 13340.

\bibitem[Scholten {\em et~al.\/}(2008)Scholten, van~der Linden \&
  This]{ScholtenLindenThis2008}
{\sc Scholten, Elke, van~der Linden, Erik \& This, Herve} 2008 The life of an
  anise-flavored alcoholic beverage: Does its stability cloud or confirm
  theory? {\em Langmuir\/} {\bf 24}, 1701--1706.

\bibitem[Sengers(1985)]{Sengers1985}
{\sc Sengers, Jan~V.} 1985 Transport properties of fluids near critical points.
  {\em Int. J. Thermophys.\/} {\bf 6}, 203.

\bibitem[Slezov(2009)]{Slezov2009}
{\sc Slezov, Vitaly~V.} 2009 {\em Kinetics of First-Order Phase Transitions\/}.
  Weinheim: Wiley-VCH.

\bibitem[Soltzberg {\em et~al.\/}(1997)Soltzberg, Bowers \&
  Hofstetter]{SoltzbergBowersHofstetter1997}
{\sc Soltzberg, Leonard~J., Bowers, Peter~G. \& Hofstetter, Christine} 1997 A
  computer model for soda bottle oscillations: {``The Bottelator''}. {\em J.
  Chem. Edu.\/} {\bf 74}~(6), 711 -- 714.

\bibitem[Sparks {\em et~al.\/}(1993)Sparks, Huppert, Kozaguchi \&
  Hallwood]{SparksHuppertKozaguchiHallwood1993}
{\sc Sparks, R.~Stephen, Huppert, Herbert~E., Kozaguchi, Takehiro \& Hallwood,
  Mark~A.} 1993 Origin of modal and rhythmic igneous layering by sedimentation
  in a convecting magma chamber. {\em Nature\/} {\bf 361}, 246--249.

\bibitem[Steinhoff \& Woermann(1995)]{Steinhoff1995}
{\sc Steinhoff, B. \& Woermann, D.} 1995 Slowing down of the kinetics of
  liquid/liquid phase separation along the binodal curve of a binary liquid
  mixture with a miscibility gap approaching the critical point. {\em J. Chem.
  Phys.\/} {\bf 103}~(20), 8985--8992.

\bibitem[Stevens \& Feingold(2009)]{StevensFeingold2009}
{\sc Stevens, Bj\"orn \& Feingold, Graham} 2009 Untangling aerosol effects on
  clouds and precipitation in a buffered system. {\em Nature\/} {\bf 461},
  607--613.

\bibitem[Stevens \& Seifert(2008)]{StevensSeifert2008}
{\sc Stevens, Bj\"orn \& Seifert, Axel} 2008 Understanding macrophysical
  outcomes of microphysical choices in simulations of shallow cumulus
  convection. {\em J. Met. Soc. Japan. Ser. II\/} {\bf 86A}, 143--162.

\bibitem[Sugimoto(1992)]{Sugimoto1992}
{\sc Sugimoto, Tadao} 1992 The theory of the nucleation of monodisperse
  particles in open systems and its application to {AgBr} systems. {\em J.
  Coll. Interf. Sc.\/} {\bf 150}~(1), 208 -- 225.

\bibitem[Taylor \& Acrivos(1964)]{TaylorAcrivos1964}
{\sc Taylor, T.D. \& Acrivos, Andreas} 1964 On the deformation and drag of a
  falling viscous drop at low reynolds numbers. {\em J. Fluid Mech.\/} {\bf
  18}~(3), 466 -- 476.

\bibitem[Tokano(2011)]{Tokano2011}
{\sc Tokano, Tetsuya} 2011 Precipitation climatology on {Titan}. {\em
  Science\/} {\bf 331}, 1393 -- 1394.

\bibitem[Tokuyama \& Enomoto(1993)]{TokuyamaEnomoto1993}
{\sc Tokuyama, M. \& Enomoto, Y.} 1993 Theory of phase-separation dynamics in
  quenched binary mixtures. {\em Phys. Rev. E\/} {\bf 47}~(2), 1156.

\bibitem[Toramaru \& Maeda(2013)]{ToramaruMaeda2013}
{\sc Toramaru, Atsushi \& Maeda, Kazuki} 2013 Mass and style of eruptions in
  experimental geysers. {\em J. Volcanology Geothermal Res.\/} {\bf 257},
  227--239.

\bibitem[Vollmer {\em et~al.\/}(1997)Vollmer, Strey \& Vollmer]{vollmer97JCP1}
{\sc Vollmer, Doris, Strey, R. \& Vollmer, J\"urgen} 1997 Oscillating phase
  separation in microemulsions {I}: Experimental observation. {\em J. Chem.
  Phys.\/} {\bf 107}~(9), 3619--3626.

\bibitem[Vollmer {\em et~al.\/}(2007)Vollmer, Auernhammer \&
  Vollmer]{vollmer07PRL}
{\sc Vollmer, J\"urgen, Auernhammer, G\"unter~K. \& Vollmer, Doris} 2007
  Minimal model for phase separation under slow cooling. {\em Phys. Rev.
  Lett.\/} {\bf 98}, 115701.

\bibitem[Vollmer {\em et~al.\/}(2014)Vollmer, Papke \&
  Rohloff]{VollmerPapkeRohloff2014}
{\sc Vollmer, J\"urgen, Papke, Ariane \& Rohloff, Martin} 2014 Ripening and
  focusing of aggregate size distributions with overall volume growth. {\em
  Front. Physics\/} {\bf 2}, 18.

\bibitem[Vollmer \& Vollmer(1999)]{vollmer99}
{\sc Vollmer, J\"urgen \& Vollmer, Doris} 1999 Cascade nucleation in the phase
  separation of amphiphilic mixtures. {\em Faraday Disc.\/} {\bf 112}, 51--62.

\bibitem[Wallace \& Hobbs(2006)]{WallaceHobbsBook2006}
{\sc Wallace, John~M. \& Hobbs, Peter~V.} 2006 {\em Atmospheric Science -- An
  Introductory Survey\/}, {\em International Geophysics Series\/}, vol.~92.
  Burlington, MA: Academic Press.

\bibitem[Weast {\em et~al.\/}(1988)Weast, Astle \& Beyer]{Weast1988}
{\sc Weast, Robert~C., Astle, Melvin~J. \& Beyer, William~H.}, ed. 1988 {\em
  CRC Handbook of Chemistry and Physics\/}, 69th edn. CRC Press, Inc. Boca
  Raton, Florida.

\bibitem[Wilkinson(2014)]{Wilkinson2014}
{\sc Wilkinson, Michael} 2014 A test-tube model for rainfall. {\em EPL\/} {\bf
  106}, 40001.

\bibitem[Woods(2010)]{Woods2010}
{\sc Woods, Andrew~W.} 2010 Turbulent plumes in nature. {\em Ann. Rev. Fluid
  Mech.\/} {\bf 42}, 391--412.

\bibitem[Wylie {\em et~al.\/}(1999)Wylie, Voight \&
  Whitehead]{WylieVoightWhitehead1999}
{\sc Wylie, Jonathan~J., Voight, Barry \& Whitehead, J.~A.} 1999 Instability of
  magma flow from volatile-dependent viscosity. {\em Science\/} {\bf
  285}~(5435), 1883--1885.

\bibitem[Yuan {\em et~al.\/}(2004)Yuan, Thomas \& Vanka]{YuanThomasVanka2004}
{\sc Yuan, Quan, Thomas, Brian~G. \& Vanka, S.~P.} 2004 Study of transient flow
  and particle transport in continuous steel caster molds: Part ii. particle
  transport. {\em Metall. Mat. Trans. B\/} {\bf 35}, 703--714.

\bibitem[Zhang \& Kling(2006)]{ZhangKling2006}
{\sc Zhang, Youxue \& Kling, George} 2006 Dynamics of lake eruptions and
  possible ocean eruptions. {\em Annu. Rev. Earth Planet. Sci.\/} {\bf 34},
  293--324.

\bibitem[Zhang \& Xu(2008)]{ZhangXu2008}
{\sc Zhang, Youxue \& Xu, Zhengjiu} 2008 {``Fizzics''} of bubble growth in beer
  and champagne. {\em Elements\/} {\bf 4}~(4), 47 -- 49.

\bibitem[Zhang(1996)]{Zhang1996}
{\sc Zhang, Y.~X.} 1996 Dynamics of {CO$_2$-driven} lake eruptions. {\em
  Nature\/} {\bf 379}~(6560), 57--59.

\end{thebibliography}


\end{document}